\documentclass{article}

\usepackage{arxiv}

\usepackage[utf8]{inputenc} 
\usepackage[T1]{fontenc}    
\usepackage[hidelinks]{hyperref}       
\usepackage{url}            
\usepackage{booktabs}       
\usepackage{amsfonts}       
\usepackage{nicefrac}       
\usepackage{microtype}      
\usepackage{lipsum}		
\usepackage{graphicx}
\usepackage{natbib}
\usepackage{doi}

\usepackage{amsmath}
\usepackage{amssymb}
\usepackage{amsthm}
\usepackage{natbib}
\usepackage{fullpage}
\usepackage{xcolor}

\usepackage{amsmath}
\usepackage{wrapfig}
\usepackage{xfrac}
\usepackage{wrapfig}
\usepackage{multirow}
\usepackage{subcaption}
\usepackage{booktabs}

\usepackage{epstopdf}
\usepackage{svg}

\newcommand{\Expect}{\mathbb{E}}
\newcommand{\Reals}{\mathbb{R}}
\newcommand{\KL}{\textrm{KL}}

\DeclareMathOperator*{\argmax}{arg\,max}
\DeclareMathOperator*{\argmin}{arg\,min}

\newcommand{\rulesep}{\unskip\ \vrule\ }

\begin{document}


\title{Variational Inference for Deblending Crowded Starfields}

\author{Runjing Liu\\
	Department of Statistics \\
	University of California, Berkeley\\
	\texttt{runjing\_liu@berkeley.edu} \\
	\And
	\href{https://orcid.org/0000-0003-2626-7320}{\includegraphics[scale=0.06]{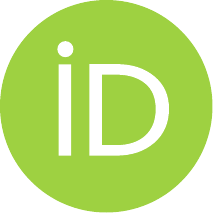}\hspace{1mm}Jon D. McAuliffe}\\
	Department of Statistics \\
	University of California, Berkeley\\
	\texttt{jon@stat.berkeley.edu} \\
	\And
        \href{https://orcid.org/0000-0002-1472-5235}{\includegraphics[scale=0.06]{orcid.pdf}\hspace{1mm}Jeffrey Regier}\\
	Department of Statistics \\
	University of Michigan\\
	\texttt{regier@umich.edu} \\
	\And \textnormal{and the} LSST Dark Energy Science Collaboration
}

\date{}

\maketitle

\begin{abstract}
In images collected by astronomical surveys, stars and galaxies often overlap visually. Deblending is the task of distinguishing and characterizing individual light sources in survey images. We propose StarNet, a Bayesian method to deblend sources in astronomical images of crowded star fields. StarNet leverages recent advances in variational inference, including amortized variational distributions and an optimization objective targeting an expectation of the forward KL divergence. In our experiments with SDSS images of the M2 globular cluster, StarNet is substantially more accurate than two competing methods: Probabilistic Cataloging (PCAT), a method that uses MCMC for inference, and DAOPHOT, a software pipeline employed by SDSS for deblending. In addition, the amortized approach to inference gives StarNet the scaling characteristics necessary to perform Bayesian inference on modern astronomical surveys.
\end{abstract}

\section{Introduction}
\label{sec:intro}

Astronomical images record the arrival of photons from distant light sources.
Astronomical catalogs are constructed from these images.
Catalogs label light sources as stars, galaxies, or other objects;
they also list the physical characteristics of light sources such as flux, color, and morphology.
These catalogs are the starting point for many downstream analyses.
For example, Bayestar used a catalog of stellar fluxes and colors to infer
the 3D distribution of interstellar dust~\citep{Green_2019_argonaut}.
Catalogs of galaxy morphologies have also been used to validate theoretical models of dark matter and dark energy~\citep{Abbott2018}.

A light source, be it a star or a galaxy, produces a peak brightness intensity at some location in an image.
When light sources are well separated, catalog construction is relatively straightforward:
characteristics of each light source, such as flux, can be estimated by analyzing
intensities at the peak and surrounding pixels.
However, in images crowded with many light sources,
observed pixel intensities may result from the combined light of multiple sources.
Source separation, or {\itshape deblending}, is the task of differentiating and
characterizing individual light sources from a mixture of intensities in an image.
A key challenge in deblending is inferring whether an observed intensity is blended,
that is, whether it is composed of a single source or multiple (dimmer) sources.


Deblending is challenging for several reasons.
First, it is an unsupervised problem without ground truth labeled data.
Second, it is a problem with a sample size of one: there is only one night sky, which is imaged many times, and the collected survey images capture overlapping regions of it.
Third, for blended fields, the properties of light sources are ambiguous; therefore, providing calibrated uncertainties for catalog construction is as important as making accurate predictions.
Finally, the scale of the data is immense.
The upcoming Rubin Observatory Legacy Survey of Space and Time (LSST),
scheduled to begin full operations by 2024,
is expected to produce 60 petabytes of astronomical images over its lifetime~\citep{LSSTabout}.

As more powerful telescopes are developed, and their ability to detect more distant light sources improves,
the density of the imaged light sources will increase.
For instance, \cite{bosch2018hyper} estimates that 58\% of light sources are blended in images captured by the
Subaru Telescope’s Hyper Suprime-Cam, and that percentage is expected to increase for LSST~\citep{sanchez2021effects}.
Therefore, developing a method that reliably characterizes light sources, even in cases of significant blending,
advances astronomical research that derives conclusions about the physical universe from estimated catalogs.

We focus on cataloging applications in which all light sources are well modeled as points without spatial extent.
Point-source-only models are applicable to surveys such as the Dark Energy Camera (DECam) plane survey,
which imaged the center of the  Milky Way \citep{Schlafly_2018_DECam}, and
the Wide-field Infrared Survey Explorer,
which has a  telescope resolution that does not allow for differentiation between stars and galaxies \citep{Wright_2010_WISESurvey}.
We use images from the Sloan Digital Sky Survey (SDSS) of the globular cluster M2, which is a region that is densely populated with stars,
as a test bed for assessing the accuracy of our approach.
We also demonstrate the ability of our method to scale to large, modern astronomical surveys
by cataloging a subregion of the DECam survey.

\bigbreak

\subsection{From Software Pipelines to Probabilistic Cataloging}

Traditionally, most cataloging has been performed using software pipelines.
These pipelines are algorithms that usually involve the following stages: locating the brightest peaks, estimating fluxes, and subtracting the estimated light sources.
These stages may be performed iteratively.
Pipelines do not normally produce statistically calibrated error estimates that propagate
the uncertainty accumulating in each of the steps.
Failure to properly accumulate error at each step results in unreliable point estimates
for images in which ambiguity exists in identifying sources and estimating their characteristics.
For example, PHOTO~\citep{lupton2001sdss}, the default cataloging pipeline used by SDSS, failed to produce a catalog of the Messier 2 (M2), a globular cluster ~\citep{Portillo_2017}.

In contrast, {\itshape probabilistic} cataloging posits a statistical model consisting of a likelihood for the observed image given a catalog and a prior distribution over possible catalogs~\citep{Portillo_2017, Brewer_2013, Feder_2019}.
Instead of deriving a single catalog, probabilistic cataloging produces a posterior distribution over the set of all possible catalogs.
Uncertainties are quantified by the posterior distribution.
For example, in an image with an ambiguously blended bright peak, some catalogs sampled from the posterior would contain multiple dim light sources while others would contain one bright source.
The relative density that the posterior distribution places on one explanation
relative to others another represents the statistical confidence in that explanation.
Moreover, a distribution over the set of all catalogs induces a distribution on any estimate derived from a catalog. Therefore, calibrated uncertainties can be propagated to downstream analyses.

Previous work on probabilistic cataloging employed Markov chain Monte Carlo (MCMC) to sample from the posterior distribution.
The MCMC procedure in~\cite{Portillo_2017} and \cite{Feder_2019}
is called PCAT, short for ``Probabilistic CATaloging."\footnote{
We use ``probabilistic cataloging'' to refer to any method that produces a posterior over possible catalogs, whereas ``PCAT" refers specifically to the MCMC procedure in~\cite{Portillo_2017} and \cite{Feder_2019}. }
A difficulty in any probabilistic cataloging approach is that the number of sources in a catalog is unknown and random, so the latent variable space is transdimensional. PCAT sampled transdimensional catalogs with reversible jump MCMC~\citep{Green95reversiblejump}, in which
auxiliary variables are added to encode the ``birth" and ``death" of light sources in the Markov chain.

The computational cost of MCMC for this model is problematic for large-scale astronomical surveys.
Early implementations of PCAT required a day to process a $100\times 100$-pixel SDSS image of M2~\citep{Portillo_2017}.
More recent implementations running inexact MCMC brought the runtime down to 30  minutes~\citep{Feder_2019}.
However, a $100\times 100$ pixel image covers only a $0.66 \times 0.66$ arcminute patch of the sky.
For comparison, in one night, SDSS scans a region on the order of $100 \times 1000$ arcminutes.
Extrapolating the 30-minute runtime suggests that PCAT would take on the order of ten years to process a nightly SDSS run.
The LSST survey will be even larger, collecting five trillion pixels nightly~\citep{LSSTnumbers}, which would require $28,000$ years to catalog using PCAT. 

As an alternative to MCMC, \cite{regier2019_celeste} proposed to use variational inference (VI)
to approximate the posterior.
VI considers a family of candidate approximate posteriors and employs numerical optimization to find the distribution in the family closest
in KL divergence to the exact posterior~\citep{Jordan_intro_vi, Wainwrite_graph_models_vi, Blei_2017_vi_review}.
With a sufficiently constrained family of distributions, the VI optimization problem can be solved orders of magnitude faster than MCMC runs.

However, \cite{regier2019_celeste} is limited in that the number of light sources in a given image is treated as known and fixed---it had to be set using a preprocessing routine.
The authors avoided the transdimensional latent variable space induced by
the unknown number of sources
in order to have a tractable objective for numerical optimization.

\bigbreak

\subsection{Our Contribution}

\nopagebreak[4]

We propose {\itshape StarNet}, an approach to deblending that employs several recent VI innovations~\citep{zhang2019advances,le2020revisiting}.
Unlike \cite{regier2019_celeste}, our VI approach is able to handle arbitrary probabilistic models, including a transdimensional model with an unknown number of sources. Section~\ref{sec:gen_model} introduces the statistical model, which is similar to the model used in PCAT.

Secondly, again unlike~\cite{regier2019_celeste},
we employ amortization, which enables StarNet to scale inference to large astronomical surveys.
In amortized variational inference, a neural network maps input images to an approximate posterior over catalogs.
Following a one-time cost to fit the neural network, inference
on new images requires just a single neural-network evaluation.
Rapid inference is possible without the need to re-run MCMC or numerically optimize VI for each new image.
For StarNet, a network evaluation, or ``forward pass," on
a $100 \times 100$ pixel image takes less than a second (vs.~30 minutes for inference using PCAT).
Section~\ref{sec:var_inference} details the variational distribution and neural network architecture in StarNet.

Finally, and critically, StarNet is fit using an expected
``forward" Kullback–Leibler (KL) divergence
between the approximate posterior $q$ and the exact posterior $p$,
where the expectation is taken over the data distribution defined by the statistical model. 
In contrast, traditional variational inference minimizes the ``reverse"
KL divergence~\citep{bishop2006pattern}, 
which uses an expectation with respect to the variational distribution. 
The forward KL is minimized using stochastic gradient descent (SGD),
which involves sampling complete data---images and their corresponding catalogs---from
their joint likelihood and fitting the network in a supervised fashion.
Section~\ref{sec:wake_sleep} details our inference procedure.

In this application, optimizing the forward KL produces more reliable approximate
posteriors than optimizing the traditional reverse KL (Section \ref{sec:elbo_sleep_compare}):
taking advantage of complete data allows the network to better avoid shallow local minima
where the approximate posterior is far from the exact posterior in terms of KL divergence.



The forward KL has been used in previous research to train deep generative
models~\citep{ambrogioni2019favi,le2020revisiting},
and appears in the sleep phase of the wake-sleep algorithm
\citep{Hinton1995wake_sleep, bornschein2014reweighted,le2020revisiting}. 
Variational inference using the forward KL is an example of simulation-based inference, 
where approximate posteriors are constructed for
likelihoods from which sampling is easy, but are unavailable analytically \citep{papamakarios2016,greenberg2019}. 
Simulation-based inference has found applications in physics where theory can provide realistic simulations \citep{Cranmer_2020}.
For example,~\cite{Baydin2019} use simulation-based inference to model
time-series data of particle paths at the Large Hadron Collider, 
and they use the forward KL objective to fit a recurrent neural network, 
whose output are proposals to an MCMC sampling scheme. 
To the best of our knowledge, our work is the first to combine amortized inference 
with the forward KL divergence to perform Bayesian inference over a transdimensional 
latent space, 
producing an approximate posterior distribution over sets. 


We applied StarNet to an SDSS image of M2, a globular cluster (Section~\ref{sec:results_on_m2})
and show that
StarNet was more accurate than the MCMC-based cataloger PCAT: though MCMC is asymptotically exact, it often suffers from incomplete mixing on practical timescales.
StarNet was also more accurate that traditional deterministic cataloging approaches in several metrics.
We then demonstrate the scalability of StarNet by cataloging a DECam image of the Milky Way (Section~\ref{sec:results_on_decam}).
Our approximate Bayesian method can produce scientifically relevant results on the order of minutes, while running PCAT would take on the order of days.


Code to reproduce our results is publicly available in a GitHub repository~\citep{BLISS2023}.

\section{The Generative Model}
\label{sec:gen_model}

In crowded starfields such as globular clusters and the galactic plane of the Milky Way, the vast majority of light sources are stars.
An astronomical image records the number of photons that reached a telescope and arrived at each pixel.
Typically, photons must pass through one of several filters, each selecting photons from a specified band of wavelengths, before being recorded.

For a given $H \times W$ pixel image with $B$ filter bands, our goal is to infer a catalog of
stars.
The catalog specifies the number of stars
in an image; for each such star, the catalog
records its location and its flux (brightness)
in each band.
The space of latent variables
$\mathcal{Z}$ is the collection of all possible catalogs of the form
\[z := \{N, (\ell_i, f_{i,1}, ..., f_{i,B})_{i = 1}^N\},\]
where the number of stars in the catalog
is $N\in\mathbb{N}$;
the location of star $i$ is $\ell_i \in \Reals^2$; and
the flux of the star $i$ in band $b$ is $f_{i, b}\in\Reals^+$.

A Bayesian approach requires specification of a prior over catalog space $\mathcal{Z}$ and a likelihood for the observed images. Our likelihood and prior, detailed below, are similar to previous approaches~\citep{Brewer_2013, Portillo_2017, Feder_2019}, 
which facilitates the comparisons of inference algorithms in isolation of model differences. 

\subsection{The Prior}
The prior over $\mathcal{Z}$ is a marked spatial Poisson process. To sample the prior, first draw the number of stars contained in the $H\times W$ image as
\begin{align}
	N &\sim \text{Poisson}(\mu HW),
	\label{eq:n_prior}
\end{align}
where $\mu$ is a hyperparameter specifying the average number of sources per pixel.
Next, draw locations
\begin{align}
  \ell_1, ..., \ell_N \mid N &\stackrel{iid}{\sim} \text{Uniform}([0, H] \times [0, W]).
 \end{align}
The fluxes in the first band are from a power law distribution with slope $\alpha$:
\begin{align}
    f_{1, 1}, ..., f_{N,1} \mid N &
    \stackrel{iid}{\sim} \text{Pareto}(f_{min}, \alpha)
    \label{eq:flux_prior}.
\end{align}
Fluxes in other bands are described relative to the first band. Like~\cite{Feder_2019}, we define the log-ratio of flux relative to the first band as ``color." Colors are drawn from a Gaussian distribution
\begin{align}
  c_{1, b}, ..., c_{N,b} \mid N  &
      \stackrel{iid}{\sim} \mathcal{N}(\mu_c, \sigma^2_c), \quad b = 2, ..., B.
\end{align}
Given the flux in the first band $f_{i,1}$ and color $c_{i,b}$,
the flux in band $b$ is  $f_{i,b} = f_{i,1} \times 10^{c_{i,b} / 2.5}$.

We set the power law slope $\alpha = 0.5$ and use a standard Gaussian for the color prior ($\mu_c = 0$, $\sigma^2_c = 1$), as in \cite{Feder_2019}.

Rather than having a hierarchical structure, 
the prior parameters are fixed in this model:
our goal is to produce a posterior on catalogs for a
specific image, not to model the population over many images. 
Appendix~\ref{sec:prior_sensitivity} evaluates the
sensitivity of the resulting catalog to choices of the prior parameters.

\subsection{The Likelihood}
Let $x_{hw}^b$ denote the observed number of photoelectrons at pixel $(h,w)$ in band $b$.
For each band, at every pixel, the expected number of photoelectron arrivals is $\lambda^b_{hw}(z)$, a deterministic function of the catalog $z$. Motivated by the Poissonian nature of photon arrivals and
the large photon arrival rate in SDSS and LSST images,
observed pixel intensities are drawn as
\begin{align}
  x_{hw}^b \mid z \overset{ind}{\sim} \mathcal{N}(\lambda^b_{hw}, \lambda^b_{hw}),
  \quad
  b = 1, ..., B; \;
  h = 1,..., H; \;
  w = 1, ..., W, \\
 \text{where } \quad
 \lambda^b_{hw} = I_b + \sum_{i = 1}^N f_{i,b} \mathcal{P}_b\big(h - \ell_{i, 1}, w - \ell_{i, 2}\big).
  \label{eq:expected_intensity}
\end{align}
Here, $\mathcal{P}_b$ is the point spread function (PSF) for band $b$ and $I_b$ is the background intensity.
The PSF is a function
$\mathcal{P}_b : \Reals \times \Reals \mapsto \Reals^+$,
describing the appearance of a stellar point source at any 2D position of the image.
Our PSF model is a weighted average between a Gaussian ``core" and a power-law ``wing" as described in~\cite{Xin2018psf}. For each band, the PSF has the form

\begin{align}
    \mathcal{P}(u,v) =
    \frac{\exp(\frac{-(u^2 + v^2)}{2\sigma_1^2}) +
    \zeta \exp(\frac{-(u^2 + v^2)}{2\sigma_2^2}) +
    \rho(1 + \frac{v^2 + u^2}{\gamma\sigma^2_P})^{-\gamma/2} }{1 + \zeta + \rho}.
\end{align}
The PSF parameters are allowed to vary by band.
In our applications to SDSS and DECam data, we use estimates of the background
and PSF obtained from a pre-processing pipeline that are distributed by these surveys along with the images.


\section{The Variational Distribution}
\label{sec:var_inference}

The central quantity in Bayesian statistics is the posterior distribution $p(z\mid x)$.
However, in many
nontrivial probabilistic models, including our own, the posterior distribution is intractable to calculate---it requires us to compute the marginal likelihood, $p(x)$, which involves integrating over the latent variable $z$.
In our model, the latent variable space is high dimensional: it is the set of all catalogs. Approximate methods such as MCMC and variational inference are therefore required.

Variational inference~\citep{Jordan_intro_vi, Wainwrite_graph_models_vi, Blei_2017_vi_review} posits a family of distributions $\mathcal{Q}$ and seeks
the distribution $q^*\in \mathcal{Q}$ that is ``closest" to the exact posterior in $\KL$ divergence.
The defined divergence and the family $\mathcal{Q}$ are chosen such that minimizer $q^*$ will
not be too difficult to find via optimization.
We index the distributions in $\mathcal{Q}$ using a real-valued vector $\eta$, in which
case solving for the optimal variational distribution $q_{\eta^*}$
becomes a numerical optimization problem.

%
Most commonly, variational inference minimizes the ``reverse" KL divergence
between $q$ and $p$:
\begin{align}
   \eta^* &= \argmin_{\eta} \mathrm{KL}\Big[\,q_\eta(z \mid x)\, \| \,p(z \mid x)\,\Big].
   \label{eq:kl_objective}
\end{align}
Minimizing the $\KL$ divergence in~\eqref{eq:kl_objective} is equivalent to maximizing the evidence lower bound (ELBO):
\begin{align}
    \mathcal{L}_{elbo}(\eta) =
    \Expect_{q_\eta(z \mid x)}\Big[\log p(x, z) - \log q_\eta(z \mid x)\Big].
    \label{eq:elbo}
\end{align}
This equality is shown in \cite{Blei_2017_vi_review}. Computing the ELBO does not require computing the marginal distribution $p(x)$, which is intractable, or the posterior distribution $p(z \mid x)$, which would be circular.
In Section~\ref{sec:wake_sleep}, we
consider an alternative objective to \eqref{eq:kl_objective},
where we instead minimize an \textit{expectation}
of the KL divergence with its arguments $q$ and $p$ reversed.

\subsection{Amortized Variational Inference}
We describe the construction of the family $\mathcal{Q}$. 
Traditionally in variational inference, the posterior approximation $q_\eta$ depends on the data $x$ only implicitly,
in that $\eta^*$ is chosen according to~\eqref{eq:kl_objective}.
In this case, $q_\eta(z \mid x)$ is usually written $q_\eta(z)$, suppressing the dependence on $x$.
When a new data point $x^{new}$ arrives, finding a variational  approximation to the posterior $p(z^{new} \mid x^{new})$ requires solving~\eqref{eq:kl_objective} with $x = x^{new}$ through an iterative optimization procedure, which may be computationally expensive.


On the other hand,
in {\itshape amortized} variational
inference~\citep{kingma2013autoencoding, rezende2014stochastic}, $q_\eta$ is an explicit function of the data.
In our case, this means a flexible, parameterized function maps input $x$, an observed image, to a real-valued vector characterizing a distribution on the latent space $\mathcal{Z}$.
Typically, the function is a neural network, in which case the variational parameters $\eta$ are the neural network weights.
After the neural network is fitted using a collection of observed $x$'s, the approximate posterior $q_\eta(z^{new} \mid x^{new})$ for a new data point
$x^{new}$ can be evaluated with a single forward pass through the neural network.
No additional run of an optimization routine is needed for a new data point $x^{new}$.

The following subsections detail the construction of our variational distribution, which will 
be fitted in an amortized fashion.

\subsection{The Factorization}
\label{sec:factorization}


\begin{figure}[tb]
    \centering
    \includegraphics[width = 0.25\textwidth]{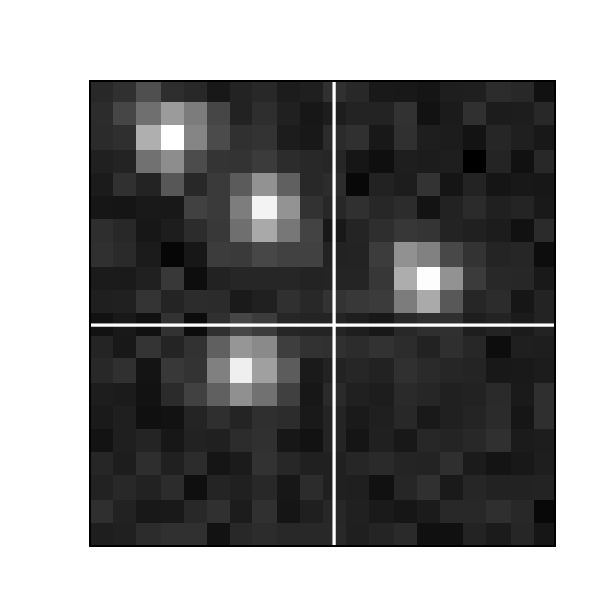}
    \caption{Tiling a $20 \times 20$ pixel image into four $10 \times 10$ tiles.}
    \label{fig:ex_tiles}
\end{figure}

To make optimization tractable, the family $\mathcal{Q}$ is normally restricted to probability distributions
without conditional dependencies between some latent variables. In the most extreme case, known as mean-field variational inference, the variational distribution completely factorizes across all latent variables.

Our factorization has a spatial structure.
First, we partition the full $H \times W$-pixel image into disjoint $R \times R$-pixel tiles.
$R$ is chosen such that the probability of having three or more stars in one tile is small.
In this way, the cataloging problem decomposes to inferring only a few stars at a time (Section~\ref{sec:nn_architecture}).

Let $S = \sfrac{H}{R}$ and $T = \sfrac{W}{R}$ and assume without loss of generality that $H$ and $W$ are multiples of $R$.
For $s = 1, ..., S$ and $t = 1, ..., T$,
the tile $\tilde x_{st}$ is composed of the pixels
\begin{align}
    \tilde x_{st} = \{x_{hw} : Rs \leq h < R(s+1) \text{ and } Rt \leq w < R(t+1)\}.
    \label{eq:tiles}
\end{align}
Figure~\ref{fig:ex_tiles} gives an example with $R = 2$.

Let $\tilde N^{(s, t)}$ be the number of stars in tile $(s,t)$.
Because $\tilde N^{(s, t)}$ is random,
the cardinality of the set of locations and fluxes in each tile
is also random.
To handle the trans-dimensional parameter space,
we consider {\itshape triangular arrays} of latent variables
for each tile:
\begin{align}
    \tilde\ell^{(s, t)} &= (\tilde\ell_{N, i}^{(s, t)} : i = 1, ..., N; N = 1, 2, ...), \\
    \text{ and } \tilde f^{(s, t)} &= (\tilde f_{N, i}^{(s, t)} : i = 1, ..., N; N = 1, 2, ...),
\end{align}
where $\tilde\ell_{N, i}^{(s, t)}$ and $\tilde f_{N, i}^{(s, t)}$ are the elements of the triangular array corresponding to location and fluxes, respectively.
Tile locations $\tilde\ell_{N, i}^{(s, t)} \in [0, R]\times[0, R]$ give the location of stars within a tile. The fluxes $\tilde f_{N, i}^{(s, t)}$ are vectors in $\mathbb{R}^B_+$ (one flux for each band).

We refer to $(\tilde N^{(s, t)}, \tilde \ell^{(s, t)}, \tilde f^{(s, t)})_{s=1,t=1}^{S,T}$ as the {\itshape tile latent variables}. The distribution on tile latent variables factorize over image tiles:
\begin{align}
    \tilde q_\eta\big( \big(\tilde N^{(s, t)}, \tilde \ell^{(s, t)}, \tilde f^{(s, t)}\big)_{s=1, t = 1}^{S, T} \mid x\big)
    &=
    \prod_{s = 1}^S \prod_{t=1}^T
    \tilde q_\eta\big(\tilde N^{(s, t)}, \tilde \ell^{(s, t)}, \tilde f^{(s, t)} \mid x\big).
    \label{eq:factorize_patches}
\end{align}

\begin{figure}[tb]
    \centering
    \includegraphics[width = 0.9\textwidth]{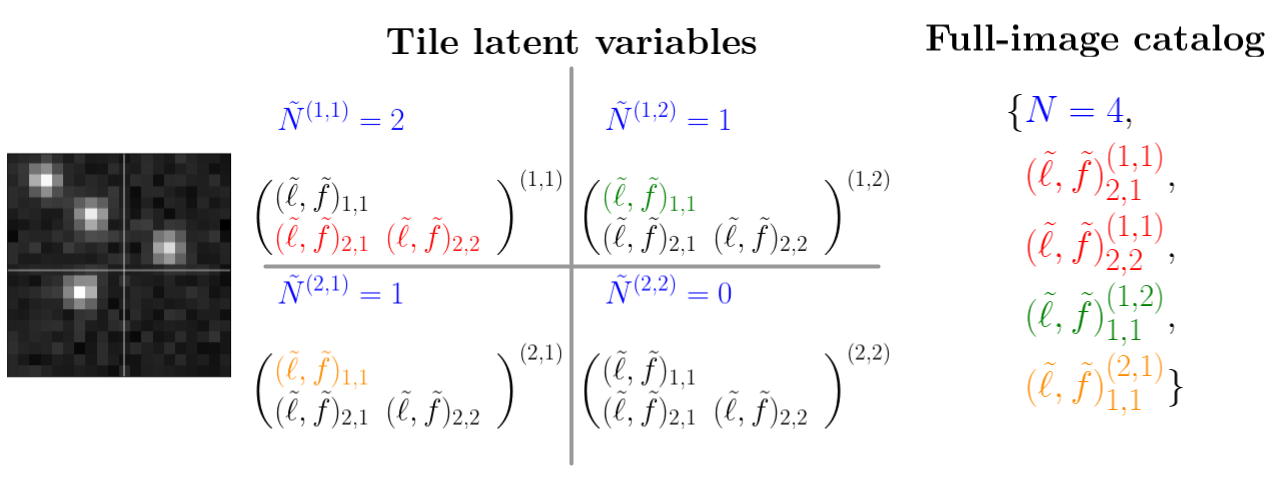}
    \caption{An example image with four tiles and four stars illustrating the relationship between the tile latent variables and the full-image catalog.
    To construct the full-image catalog, we index into the appropriate row of the triangular array for each tile.}
    \label{fig:tile_to_full_schm}
\end{figure}

We denote tile latent variables as $\tilde z$.
The ultimate latent variable of interest is $z = \{N, (\ell_i, f_{i,1}, ..., f_{i,B})_{i = 1}^N\}$, the catalog for the full image.
There is a mapping from $\tilde z$ to $z$.
First, the number of stars in the full catalog is given by the sum of the stars in each tile, $N = \sum_{s,t} \tilde N^{(s, t)}$.
Then, for every tile $(s,t)$, we index into the $\tilde N^{(s,t)}$-th row of the triangular array of tile latent variables $\tilde f^{(s,t)}$ and $\tilde \ell^{(s,t)}$.
The union of these fluxes and locations over all tiles form the full catalog (tile locations are shifted by the position of the tile in the full image to obtain locations in the full image).
See Figure~\ref{fig:tile_to_full_schm} for a schematic.




If $\tau$ is the mapping from $\tilde z$ to $z$,
then the variational distribution on catalogs $z$ is
\begin{align}
    q_\eta(z \mid x) := \tilde q_\eta(\tau^{-1}(z) \mid x),
    \label{eq:pull_back_of_q}
\end{align}
where $\tau^{-1}(z)$ is the pre-image of $z$ under $\tau$.
See Appendix~\ref{sec:eval_var_distr} for details on evaluating $q_\eta(z \mid x)$ for any given catalog $z$, which by \eqref{eq:pull_back_of_q} requires finding the pre-image $\tau^{-1}(z)$.






\subsection{Variational Distributions on Image Tiles}
\label{sec:distr_on_tiles}
We describe the variational distribution for each tile,
$\tilde q_\eta\big(\tilde N^{(s, t)}, \tilde \ell^{(s, t)}, \tilde f^{(s, t)} \mid x\big)$.
The latent variables fully factorize within each tile.
Dropping the index
$(s,t)$ in this subsection,
\begin{align}
    \tilde N &\sim \text{Categorical}(
    \omega; 0, ..., N_{max});  \label{eq:var_distr_n}\\
	\tilde \ell_{j, i} / R &\sim \text{LogitNormal}(\mu_{\ell_{j, i}}, \text{diag}(\nu_{\ell_{j, i}}) )\label{eq:var_distr_loc}; \\
	\tilde f^b_{j, i} &\sim \text{LogNormal}(\mu_{f^b_{j, i}}, \sigma^2_{f^b_{j, i}}), \label{eq:var_distr_f}
\end{align}
independently for $i = 1, ..., j$; $j = 1, ..., N_{max}$.
Here $\omega$ is a $(\tilde N_{max} + 1)$-dimensional vector on the simplex. $\mu_{\ell_{j, i}}$ and $\nu_{\ell_{j, }}$ are two-dimensional vectors---the covariance on locations is diagonal.
Note that in the exact posterior, $\tilde N$ has support on the nonnegative integers, whereas in the variational distribution $\tilde N$ is truncated at some large $N_{max}$.

These distributions were taken to match the constraints of the latent variables: fluxes are positive and right skewed, suggesting a log-normal; locations are between zero and $R$, suggesting a scaled logit-normal.

\subsection{Neural Network Architecture}
\label{sec:nn_architecture}

In each tile, the distributional parameters in \eqref{eq:var_distr_n},
\eqref{eq:var_distr_loc}, and \eqref{eq:var_distr_f} are the output of a neural network.
The input to the neural network is an $R \times R$ tile, padded with surrounding pixels.
Padding enables the neural network to produce better predictions inside the tile.
For example, a bright source outside but in the vicinity of the tile affects the pixel values inside the tile.
Padding the tiles allows the neural network access to this information.
Thus, while the distribution on tile latent variables factorize over tiles, the neural network is able to use information from neighboring tiles in producing the distributional parameters.

The appropriate amount of padding will depend on the PSF width in the analyzed image.
To catalog the crowded starfield M2 (Section~\ref{sec:results_on_m2}),
we set $R = 2$ and padded the tile with a three-pixel-wide boundary.
In cataloging a DECam image, we use larger tiles with more padding because the width of the PSF is larger in these images. There, we set $R = 10$ and used a five-pixel-wide boundary.

\begin{figure}[!tb]
    \centering
    \includegraphics[width=\textwidth]{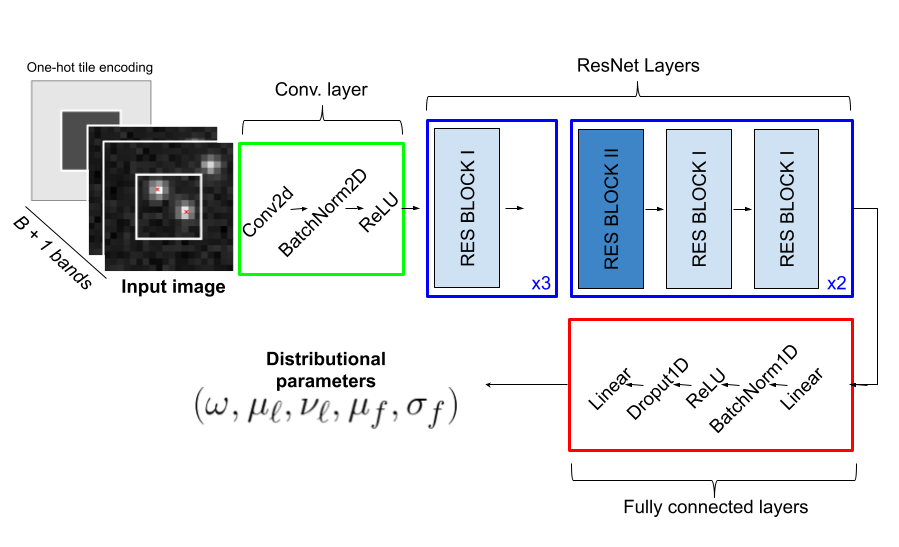}
    \caption{The neural network architecture. The DECam input is a two-band $20\times 20$ padded tile, 
    and the network returns distributional parameters corresponding to sources located in the $10\times 10$ tile outlined in white.
    In this example, there are two sources located in the $10\times10$ tile.
    The additional input band is a one-hot encoding with ones for pixels inside the tile, and zeros outside. 
    For further details concerning the residual network blocks (in blue), see 
    Appendix~\ref{sec:supp_nn_architecture}.
    }
    \label{fig:starnet_arch}
\end{figure}

In amortized inference, the variational parameters $\eta$ are neural network weights.
The architecture consists of a convolutional layer followed by several residual network layers, which themselves contain convolutions, before ending with several fully connected layers (Figure~\ref{fig:starnet_arch}).
This architecture has been successful on image classification challenges such as ImageNet~\citep{imagenet2015}.
We tuned the architecture using Optuna, an automatic hyper-parameter optimization package \citep{optuna_2019}. 
Our search included the number of convolution layers, 
the number of fully connected layers,
the number of channels in the convolution layers, and the size of the fully connected layers. 

An input to the network is a padded tile, which consists of $B$ color bands. 
We also append an additional ``band" to the input, which 
is a one-hot encoding with
ones for pixels inside the tile, and zeros outside. 
We do this because the network is only responsible for inferring sources inside each tile, and this additional band gives
the network access to a feature which encodes the tile interior. 

See Appendix~\ref{sec:supp_nn_architecture} for further details 
concerning the parameters of our neural network architecture. 


Note that the output dimension of the neural network is quadratic in $N_{max}$: the outputs are parameters for a triangular array consisting of $\frac{1}{2}(N_{max}^2 + N_{max})$ sources.
Factorizing the variational distribution spatially keeps the output dimension manageable.
While the full image may contain many stars
(on the images that we catalog, the number of stars is on the order of thousands),
we set $N_{max} = 3$ for each tile.
Thus, the network is responsible for inferring only a few stars at once---a much easier task than inferring all stars simultaneously.

We emphasize that while the variational distribution factorizes over tiles, our method does not break the inference problem for the full image into isolated subproblems.
The likelihood of the full image does not factorize over tiles. Light from a star within a tile spills over into neighboring tiles, so the likelihood should not and does not decouple across image tiles.

\section{The Expected Forward Kullback-Leibler Divergence}
\label{sec:wake_sleep}


Procedures such as black-box variational inference (BBVI)~\citep{ranganath2013black} and
automatic-differentiation variational inference (ADVI)~\citep{kucukelbir2016automatic}
optimize the ELBO  without the need for
deriving analytic expressions for the expectation over $q_\eta$.
These approaches employ stochastic gradient descent (SGD);
they sample latent variables from $q_\eta$ and produce an unbiased estimate for the gradient of the ELBO by taking advantage of modern automatic differentiation tools.
ADVI is closely related to the reparameterization trick \citep{SpallOptimization2003,kingma2013autoencoding, rezende2014stochastic}, which is often used to fit variational autoencoders and applies when the latent variables are continuous.

In our model, the number of stars $N$ is discrete.
The REINFORCE estimator~\citep{Williams1992reinforce} is one way to produce an unbiased stochastic gradient for both continuous and discrete latent variables.
However, REINFORCE gradients often suffer from high variance in practice, even with the introduction of control variates, resulting in slow convergence of stochastic optimization. We find this to be true in our empirical study (Section~\ref{sec:elbo_sleep_compare}).

The key difficulty in constructing stochastic gradients of the ELBO is that the
integrating distribution depends on the optimization parameter $\eta$.
We instead maximize a negated expectation of the ``forward" KL divergence:
\begin{align}
    \mathcal{L}_{fwd}(\eta) :=
    - \Expect_{x \sim p(x)}\Big[\KL(p(z \mid x) \| q_\eta(z \mid x))\Big]
    \label{eq:sleep_obj},
\end{align}
an objective that appears in the ``sleep phase" of the wake-sleep algorithm
\citep{Hinton1995wake_sleep, bornschein2014reweighted,le2020revisiting}
and was re-introduced by \cite{ambrogioni2019favi}, who called this approach ``Forward Amortized Variational Inference."

Section~\ref{sec:sleep_details} details a simple gradient estimator for~\eqref{eq:sleep_obj} that does not require reparameterization or REINFORCE.

There are two key differences between the expected forward KL objective~\eqref{eq:sleep_obj}
and the ELBO~\eqref{eq:elbo}.
First, recall that maximizing the ELBO is equivalent to minimizing
$\KL(q\|p)$; the KL in $\mathcal{L}_{fwd}$ transposes the arguments.
Second, the outer expectation over $p(x)$ in $\mathcal{L}_{fwd}$ gives a different meaning to the objective.
The ELBO objective seeks $\eta$ to minimize the $\KL$ between $q_\eta(z \mid x)$ and $p(z \mid x)$
for fixed, observed data $x$,
in this case the $H\times W$ image.
In contrast, minimizing $\mathcal{L}_{fwd}$ minimizes the $\KL$
on average over all possible data $x$,
as weighted by $p(x)$.
The target is no longer an approximate posterior for the observed data, but rather
an approximate posterior that is ``good on average" over all possible data under the model $p(x)$.

\subsection{Decomposing the Expected Forward KL}
\label{sec:sleep_details}
In this subsection, we decompose the expected forward KL objective $\mathcal{L}_{fwd}$ 
to obtain an unbiased stochastic gradient for stochastic gradient descent. 

First, observe that optimizing $\mathcal{L}_{fwd}$ does not require computing the intractable term $p(x)$:
\begin{align}
\argmax_\eta\; \mathcal{L}_{fwd}(\eta)
    & = \argmin_{\eta} \; \mathbb{E}_{x \sim p(x)}\Big[ \mathrm{KL}(p(z \mid x) \| q_\eta(z \mid x)\Big] \\
  &=\argmin_{\eta} \; \mathbb E_{p(x)}\Big[\mathbb E_{p(z \mid x)}\Big(\log p(z \mid x) - \log q_\eta(z \mid x) \Big)\Big]\\
&=\argmin_{\eta} \; \mathbb E_{p(x, z)}\Big[- \log q_\eta(z \mid x) \Big]\label{eq:sleep_loss_simple}.
\end{align}
Notice the integrating distribution $p(x,z)$ does not depend on the optimization parameter $\eta$.
Thus, unbiased stochastic gradients can be obtained as
\begin{align}
    g = -\nabla_\eta \log q_\eta(z \mid x) \quad \text{ for } (x, z)\sim p(x, z).
\end{align}

In other words, at each iteration of SGD, we simulate complete data $(x,z)$ from the generative model and evaluate the loss $-\log q_\eta(z \mid x)$.
Here, ``complete data" refers to the image along with its catalog.
This loss encourages the neural network to map an image $x$ to a distribution $q_{\eta}(\cdot \mid x)$ that places large density on the image's catalog $z$.

We decompose the loss $-\log q_\eta(z \mid x)$ further.
Recall that $q_\eta$ fully factorizes over tile latent variables, and thus $-\log q_\eta(z \mid x)$ is a summation over all tile latent variables.
To evaluate $-\log q_\eta(z \mid x)$ for some $(x,z)\sim p$, first convert $z$ to its tile parameterization $(\tilde N^{(s,t)}, \tilde \ell^{(s,t)}, \tilde f^{(s,t)})_{s=1,t=1}^{(S,T)}$, as detailed in Appendix~\ref{sec:eval_var_distr}.
For each tile $(s,t)$, the variational distribution on the number of stars $\tilde N^{(s,t)}$ is categorical with probability vector $\omega^{(s,t)}\in\Delta^{N_{max}}$ (recall Section~\ref{sec:distr_on_tiles}).
The loss function for the number of stars becomes
\begin{align}
    - \log q_\eta(\tilde N^{(s,t)} \mid x) = -\sum_{n = 0}^{\tilde N_{max}} 1\{\tilde N^{(s,t)} = n\} \log \omega^{(s,t)}_n.
    \label{eq:cross_entropy_loss}
\end{align}
The vector $\omega^{(s,t)}$ is the output of the neural network, and \eqref{eq:cross_entropy_loss} is the usual cross-entropy loss for a multi-class classification problem.

Next, recall that in the variational distribution location coordinates are logit-normal and fluxes are log-normal.
Let $y$ generically denote either the logit-location or log-flux for a star in the sampled catalog $z$; let $(\hat\mu, \hat\sigma^2)$ be the Gaussian mean and variance returned by the neural network. Then the loss for these latent variables is,
\begin{align}
    -\log q_\eta(y \mid x) =
        \frac{1}{2\hat\sigma^2}(y - \hat\mu)^2
         + \frac{1}{2}\log(2\pi\hat\sigma^2).
         \label{eq:gaussian_sleep_loss}
\end{align}
The first term encourages network predictions $\hat\mu$ to be close to the sampled latent variable $y$, while $\hat\sigma^2$ encodes the uncertainty of the network: the second term encourages small uncertainties, but is
balanced by the scaling of the error $(y - \hat\mu)^2$ in the first term.



The losses in~\eqref{eq:cross_entropy_loss} and~\eqref{eq:gaussian_sleep_loss} show that the
expected forward KL objective results in a supervised learning problem on complete data
sampled from our generative model:
the objective function for the number of stars is the usual cross-entropy loss for classification,
while the objective function for log-fluxes and logit-locations are $L_2$ losses in the mean parameters.




\section{Empirical Comparison of KL Objectives}


\label{sec:elbo_sleep_compare}

\begin{figure}[!htb]
    \centering
    \begin{subfigure}[t]{0.9\textwidth}
    \centering
    \includegraphics[width=\textwidth]{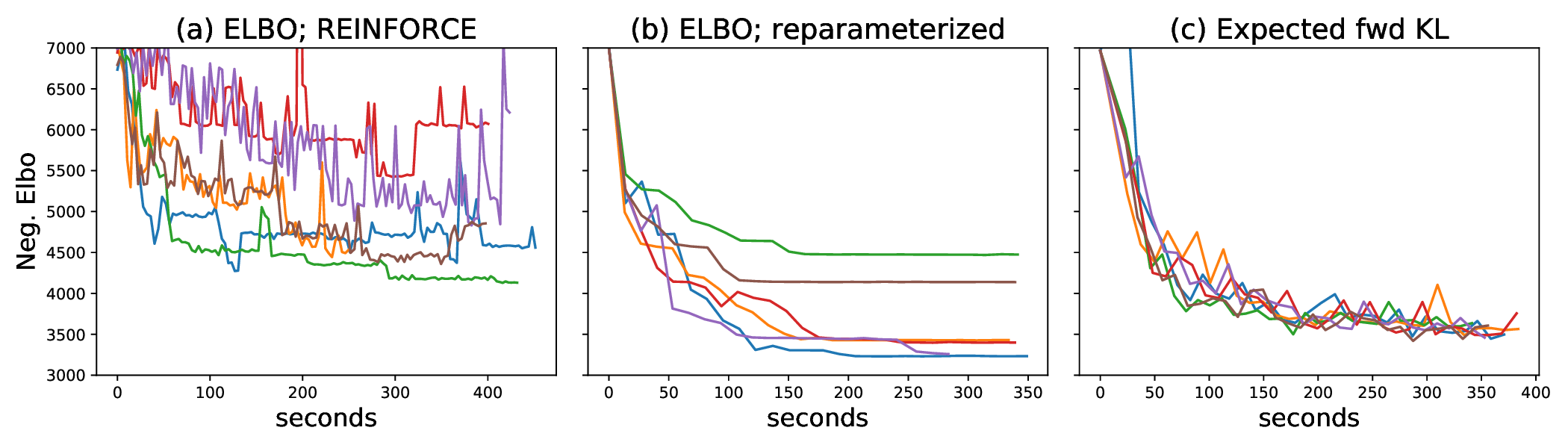}
    \end{subfigure}
    \begin{subfigure}[t]{\textwidth}
    \centering
    \includegraphics[width=0.9\textwidth]{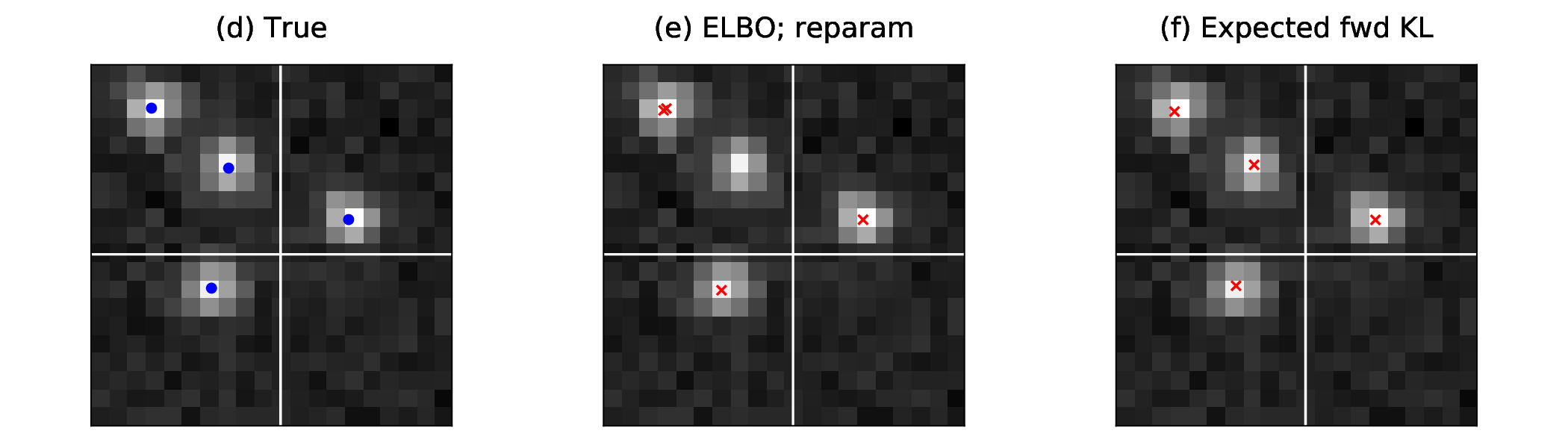}
    \end{subfigure}
    \caption{(Top row) The negative ELBO as the optimization progresses for six random restarts.
    (Bottom row) In red, modal locations from ELBO-optimized and $\mathcal{L}_{fwd}$-optimized variational posteriors, for one of the six restarts.
    In blue, the true locations. }
    \label{fig:optim_path}
\end{figure}

A simple example demonstrates that there exist shallow local optima in the ELBO
where the fitted approximate posterior is far in KL divergence from the exact posterior.
These local optima result in unreliable catalogs.
The expected forward KL, by taking advantage of complete data, appears to have a more favorable optimization landscape.
The simulated $20\times20$ single-band image $x_{test}$ is shown in Figure~\ref{fig:optim_path}(d).

We compare three approaches to deblending. The first two approaches directly optimize the ELBO,
\begin{align}
\mathcal{L}_{elbo}(\eta; x) = \Expect_{q_{\eta}(z \mid x)}\Big[\log p(x, z) - \log q_{\eta}(z \mid x)\Big],
\label{eq:elbo_on_test}
\end{align}
evaluated at $x = x_{test}$. 
The third approach minimizes the expected forward KL~\eqref{eq:sleep_obj}.
In each case, $q_\eta$ is the inference network from Section~\ref{sec:nn_architecture}.
The input to the network is a $10\times 10$ tile with no padding.

Note that the expected forward KL does not depend on $x_{test}$.
Optimizing $\mathcal{L}_{fwd}$ only requires sampling catalogs from the prior
and simulating images conditional on each catalog.
The prior on the number of stars per image was set to be Poisson with mean $\mu = 4$.

Figure~\ref{fig:optim_path}~(top row) charts the test ELBO~\eqref{eq:elbo_on_test}
as optimization proceeds in our three approaches.
The first approach optimizes the ELBO with SGD and a REINFORCE plus control-variate gradient estimator~\citep{ranganath2013black}.
The path of the ELBO objective in this first approach is irregular, likely due to the high variance of the REINFORCE gradient estimator, 
and the optimization does not appear to converge (Figure~\ref{fig:optim_path}a).
For a lower-variance gradient estimator, the second approach employed the reparameterized gradient. To employ this gradient estimator, we analytically integrated the ELBO with respect to the number of stars $N$ to remove the discrete random variable.
See Appendix~\ref{sec:reparam_details} for details about the gradient estimators.
Using reparameterized gradients instead of REINFORCE gradients enabled the optimization to converge to stationary points (Figure~\ref{fig:optim_path}b).
However, for two randomly initialized restarts,
the optimization found local optima where the negative ELBO is notably higher than other restarts.

\begin{figure}[!htb]
    \centering
    \includegraphics[width=0.9\textwidth]{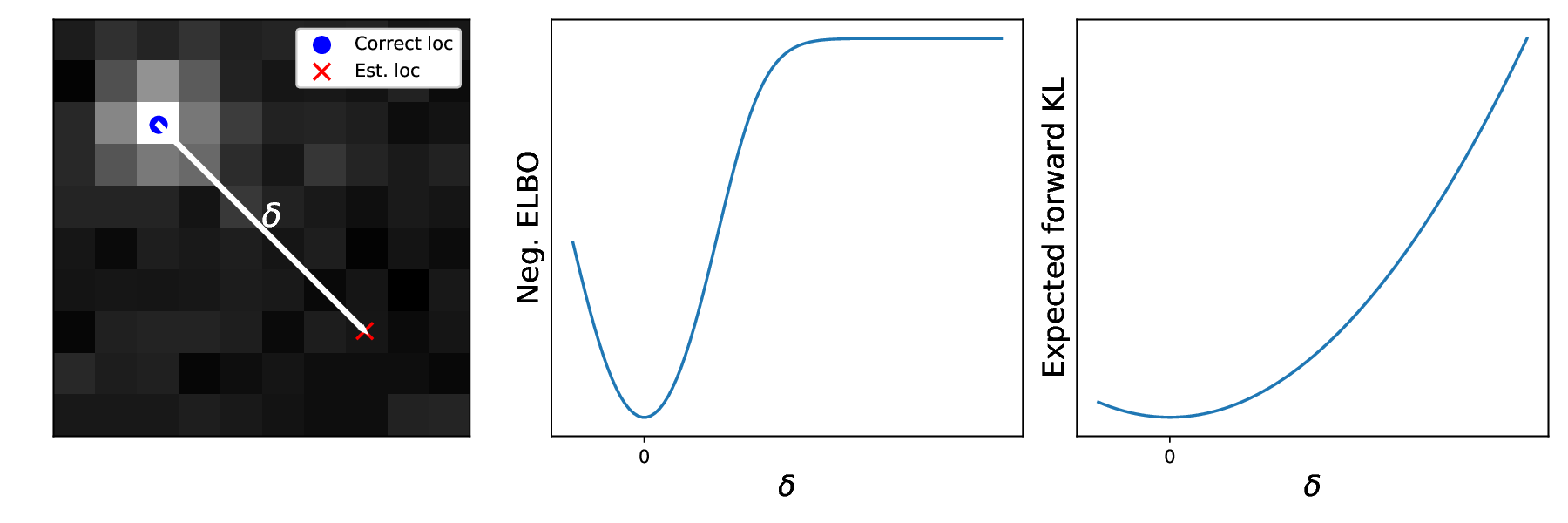}
    \caption{An illustration of local optima in the ELBO objective.
    To move an estimated location to a correct location,
    the optimization path must traverse a region where the negative ELBO is flat, with near-zero gradients.
    In contrast, when optimizing the expected forward KL with SGD, each iteration evaluates a
    quadratic loss between a true and estimated location,
    and the gradient does not vanish. }
    \label{fig:local_optima_cartoon}
\end{figure}

These shallow local optima in the ELBO result in unreliable catalogs.
The bottom row of Figure~\ref{fig:optim_path} displays the estimated locations, defined as the mode of the fitted variational distribution.
Figure~\ref{fig:optim_path}(e) shows these locations after converging to a shallow local optimum.
Here, the upper left tile was correctly estimated to have two stars, but both estimated stars were incorrectly placed at the same location.
(One of the locations should be placed on the second star.)
To move one of the estimated locations to the second star, the optimization path must traverse a region where the log-likelihood is lower than the current configuration (Figure~\ref{fig:local_optima_cartoon}).
The displayed configuration is a local optimum where the gradient with respect to its locations is approximately zero.

On the other hand, using
the expected forward KL does not directly optimize the test ELBO.
However, the test ELBO increases nonetheless, because the variational posterior
better approximates the exact posterior as the optimization proceeds.
Optimizing $\mathcal{L}_{fwd}$ consistently converged to
a similar ELBO across all restarts and avoided shallow local optima (Figure~\ref{fig:optim_path}c).
At each iteration of SGD, 
the evaluated loss is quadratic in 
the logit-location estimate
$\mu_\ell$ (\ref{eq:gaussian_sleep_loss}),
and the gradient does not vanish.
By avoiding shallow local optima, the variational distribution fit with the forward KL
always placed its mode on the four true stars in our trials.
An example of successful detection by fitting with
$\mathcal{L}_{fwd}$ is shown in~Figure~\ref{fig:optim_path}(f).

Finally, note that low-variance gradients of the ELBO for this simple example were constructed by analytically integrating out $N$, 
and this was only possible because the image consisted of only four tiles. 
For each tile, the variational distribution has support over only 0, 1, or 2 stars.
Since the variational distribution factorizes over the four tiles, integrating $N$ is a summation of $3^4 = 81$ terms.
On larger images with more tiles, analytically integrating $N$ would be computationally infeasible,
and the standard reparameterization trick would not apply as it does in this small illustrative example.

\section{Results on Astronomical Surveys}

We evaluate StarNet on two distinct surveys. First, we catalog an SDSS image of the Messier 2 (M2) globular cluster. 
We evaluate the catalog quality by validating against data collected from 
the Hubble Space telescope, which we use as a ground truth. 

Then, we run StarNet on a high-resolution DECam image of the galactic plane of the Milky Way. 
We demonstrate the ability of StarNet to scale to larger astronomical surveys.

\subsection{Results on the M2 Globular Cluster}
\label{sec:results_on_m2}

\begin{figure}[tb]
    \centering
    \includegraphics[width=0.7\textwidth]{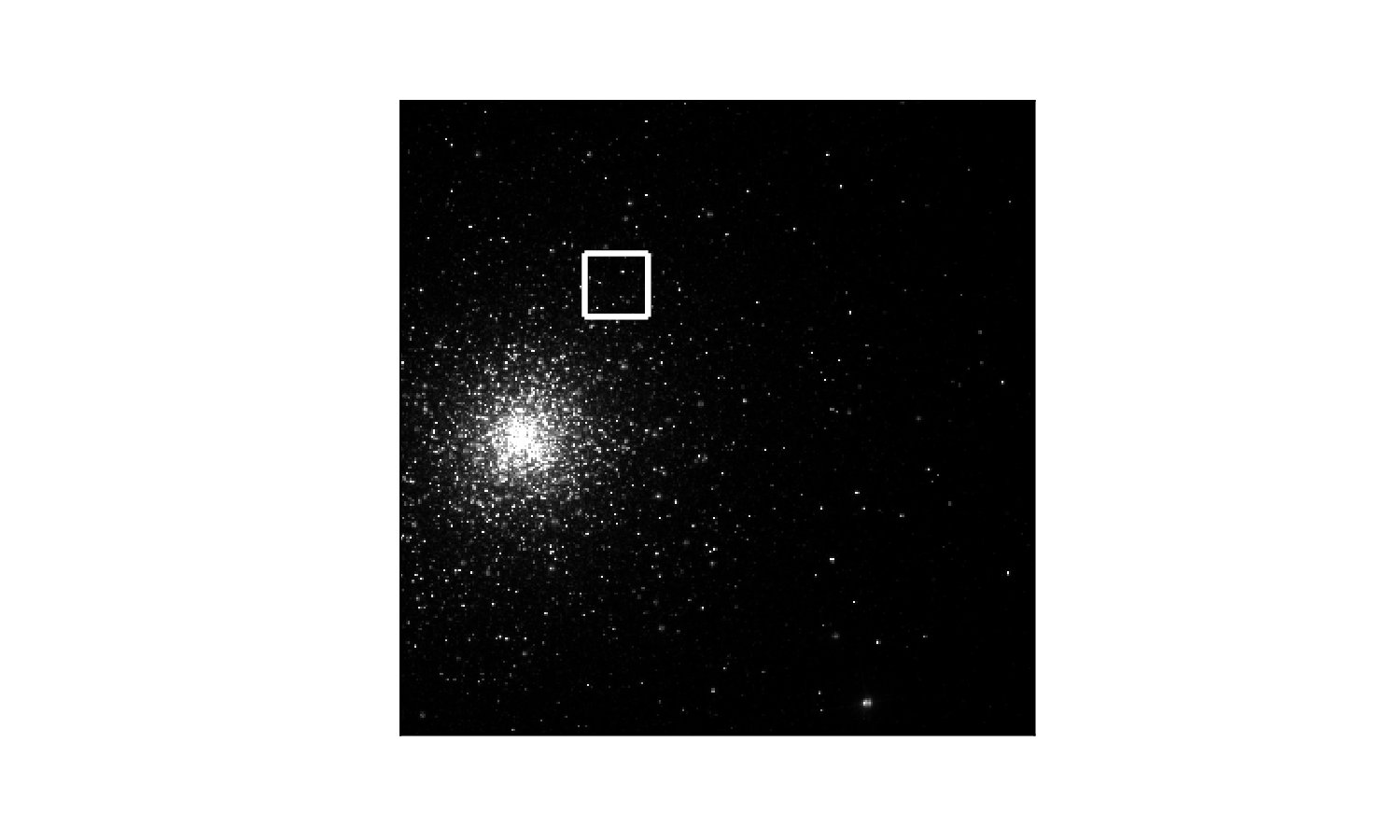}
    \caption{The M2 globular cluster as imaged by SDSS. In white is the $100 \times 100$ subregion 
    cataloged by PCAT in \cite{Feder_2019}. }
    \label{fig:m2_region}
\end{figure}

The M2 globular cluster is a crowded starfield found in field 136 of camera column 2 in run 2583 of the SDSS survey. 
M2 was also imaged in the ACS Globular Cluster Survey~\citep{Sarajedini_2007}
using the Hubble Space Telescope (HST),
which has greater resolution than the SDSS telescope.
The resolution of the HST wide-field channel is 0.05 arcseconds per pixel versus
0.40 arcseconds per pixel in SDSS \citep{hubble_about, sdss_about}.
For this image, the catalog from the HST survey (henceforth the ``HST catalog")
serves as ground truth for validating our results.

We first analyze the $100 \times 100$ pixel subimage of M2 that
\cite{Portillo_2017} and \cite{Feder_2019} analyzed with their MCMC-based approach, PCAT.
This subimage shows a region located outside the heavily saturated core of the cluster (Figure~\ref{fig:m2_region}). 
Nonetheless, in this subimage the HST catalog contains 1114 stars with F606W-band magnitudes less than 22.
We include two bands in our model, the SDSS $r$-band and $i$-band.
The SDSS $r$-band and the Hubble F606W band are centered at roughly the same wavelength,
while the wavelength range of the Hubble F606W band is slightly broader.

We compare the cataloging accuracy of StarNet
against PCAT, the aforementioned MCMC-based approach that uses the same generative model as StarNet;
and DAOPHOT, an algorithmic routine for detecting stars in crowded starfields
which does not use a probabilistic model~\citep{stetson2987daophot}.
DAOPHOT convolves the observed image with a Gaussian kernel and scans for peaks above a given threshold.
The DAOPHOT catalog of M2 was reported in \cite{An_2008_m2}.


To evaluate the three methods, we filtered the ground truth HST catalog to stars with magnitudes smaller than 22.5 in the Hubble F606W band
(note that smaller magnitude corresponds to brighter stars),
because none of the three methods were able to detect stars
with lower apparent brightness in the SDSS image.

Estimated catalogs are evaluated on three metrics: the true positive rate (TPR), or recall;
the positive predictive value (PPV), or precision;
and the F1 score.
The TPR is the proportion of true stars in the HST catalog matched with a predicted star in the estimated catalog.
The PPV is the proportion of predicted stars in the estimated catalog matched with a true star in the HST catalog.
The F1 score summarizes the two metrics as the harmonic mean of the PPV and the TPR.

Like \cite{Portillo_2017} and \cite{Feder_2019}, we define a ``match" between an estimated star
and an HST star as follows: (1) the estimated location and the HST location are within 0.5 SDSS pixels,
and (2) the estimated SDSS $r$-band flux and the HST F606W band flux are within half a magnitude.

In probabilistic cataloging (PCAT and StarNet), the posterior defines a distribution over catalogs.
For StarNet, the TPR, PPV, and F1 score were computed for the catalog corresponding to the mode of the variational distribution (henceforth, the StarNet catalog).
For PCAT, 300 catalogs were sampled using MCMC; the metrics were computed for each sampled catalog and averaged.

\begin{figure}[tb]
    \centering
    \includegraphics[width=0.49\textwidth]{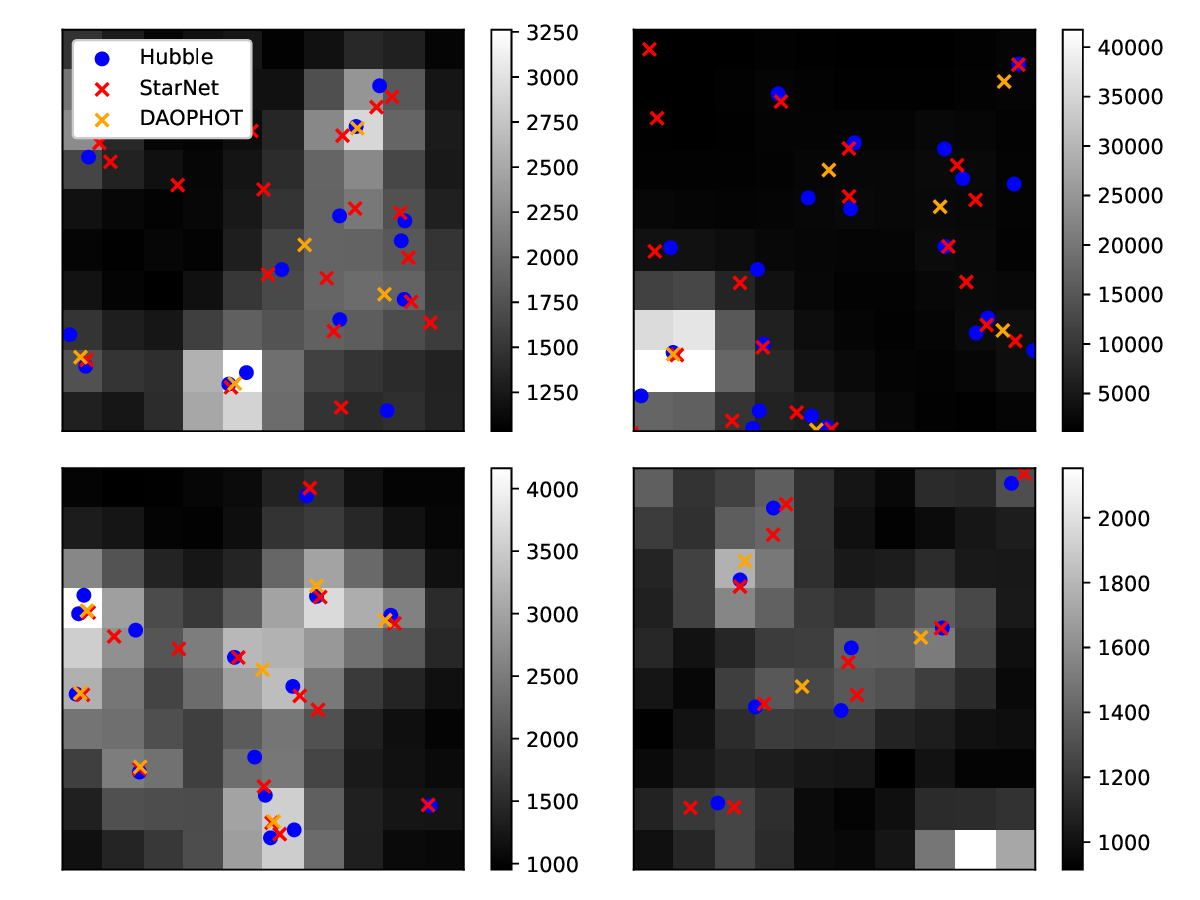}
    \rulesep
    \includegraphics[width=0.49\textwidth]{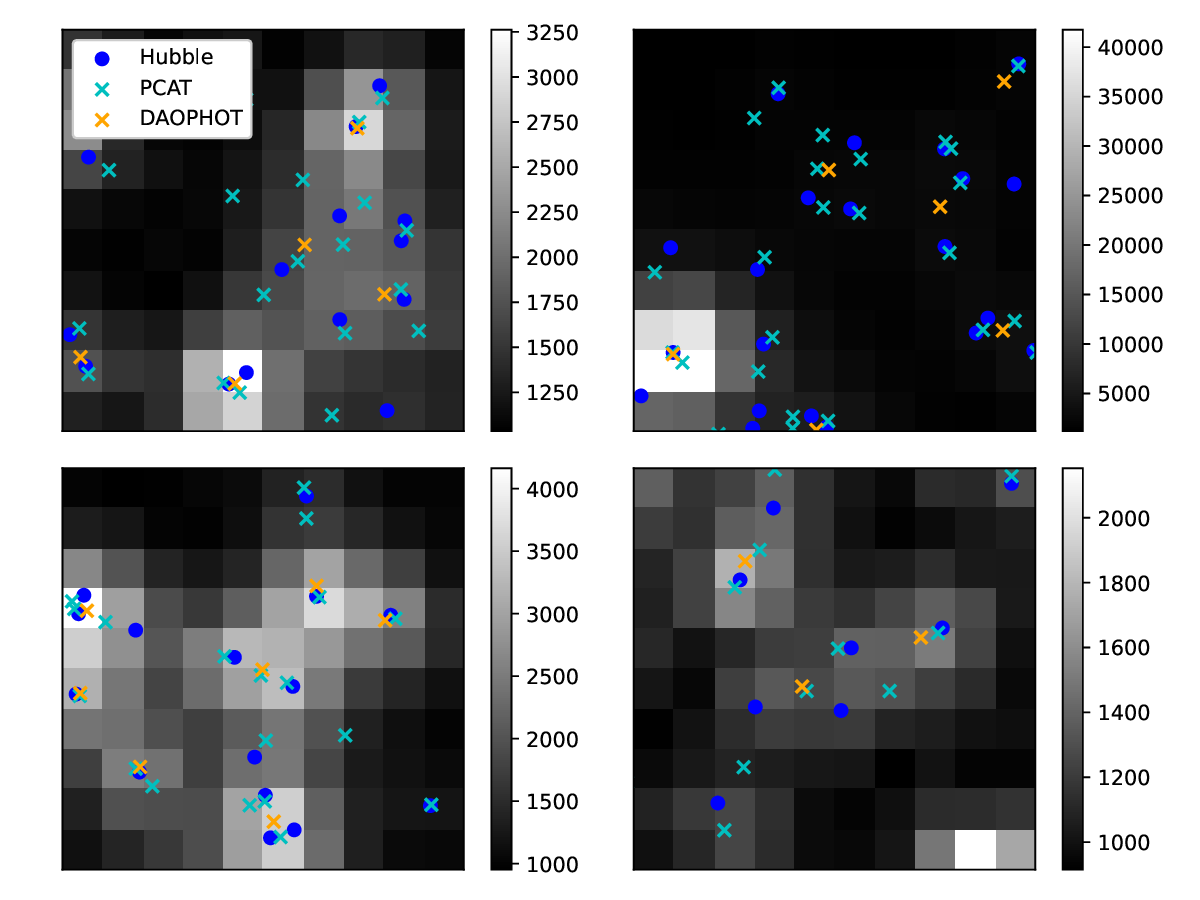}
    \caption{Estimated catalogs on four 10$\times$10 subimages from
    M2. Blue dots are stars from the HST catalog used as ground truth.
    StarNet, PCAT, and DAOPHOT estimated stars are shown as
    red, cyan, and orange crosses, respectively. }
    \label{fig:example_subimages}
\end{figure}

StarNet produced a catalog that outperforms DAOPHOT and PCAT in F1 score (Table~\ref{tab:summary_stats}).
Figure~\ref{fig:example_subimages} compares the StarNet catalog to the PCAT, DAOPHOT, and HST catalogs.
DAOPHOT estimated less than half the number of stars when compared to the other methods.
It therefore had a large PPV but a small TPR.
The StarNet catalog had similar TPR as PCAT while having an 11\% higher PPV.

\begin{figure}[tb]
    \centering
    \includegraphics[width=0.99\textwidth]{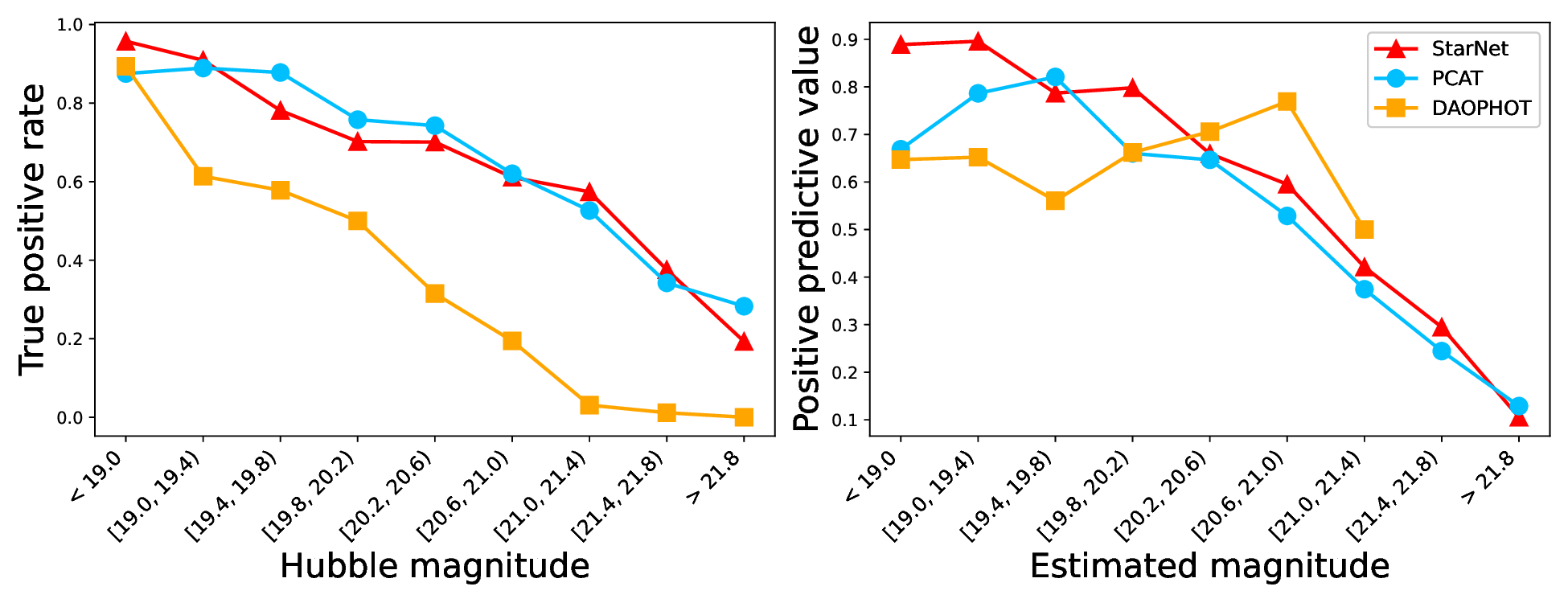}
    \caption{True positive rate (left) and positive predicted value (right) of various cataloging
    procedures on M2, plotted against $r$-band magnitude.
    Smaller magnitudes correspond to brighter stars.
    }
    \label{fig:summary_stats}
\end{figure}

\begin{figure}[tb]
    \centering
    \includegraphics[width=0.99\textwidth]{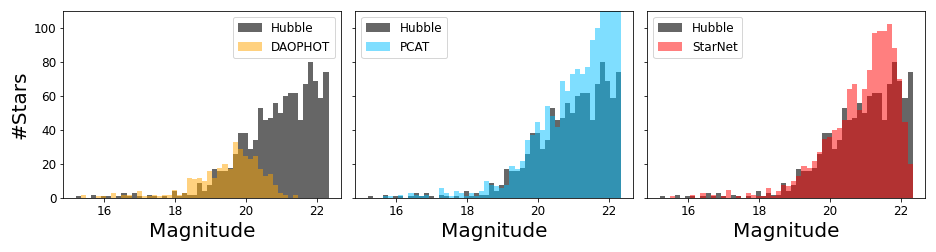}
    \caption{Flux distributions for the $r$-band observations of M2.
    The flux distribution of the HST catalog is in grey.
    Estimated distributions from DAOPHOT, PCAT, and StarNet catalogs are overlaid.
    For PCAT, the flux distribution is from a single catalog sample. }
    \label{fig:luminosity_fun_m2}
\end{figure}

The improvement of StarNet over PCAT in PPV was most pronounced for the brightest stars (Figure~\ref{fig:summary_stats}), 
suggesting that some of the brightest stars in the PCAT catalog may have in truth been collections of blended stars. 
The TPR for StarNet was uniformly better than DAOPHOT across all magnitudes.
Of all methods, StarNet best approximated the HST flux distribution (Figure~\ref{fig:luminosity_fun_m2}).

Table~\ref{tab:summary_stats} also shows the number of stars inferred by each method.
There are 1114 stars in the HST catalog.
For probabilistic methods (StarNet and PCAT),
we display the mean number of stars under the approximate posterior, along with the 5th and 95th percentiles.
We compute the StarNet posterior mean and quantiles by sampling from the variational posterior. 
Recall that on each tile, the variational posterior on the number of stars is a categorical random variable; 
to construct a distribution for the number of stars on the whole image, we first sample from the per-tile categorical distribution, then sum over all tiles. 
StarNet posterior intervals were three times wider than the PCAT intervals.
The small PCAT intervals may indicate that the MCMC sampler failed to mix well. 
While neither the StarNet intervals nor the PCAT intervals cover the ground truth, though the StarNet intervals come closer to doing so. 
For StarNet, we attribute the over-estimated number of stars by StarNet to the tiling structure of 
the approximate posterior (Appendix~\ref{sec:coverage}). 


\begin{table}[!tb]
\centering
\begin{tabular}{l|ccc|cc}
\toprule
& & & & \multicolumn{2}{c}{\#Stars} \\
     Method &   TPR &   PPV &  F1 score &  mean & (q-5\%, q-95\%)\\
\midrule
    DAOPHOT &  0.20 &  0.65 &      0.31 &     357 & -- \\
       PCAT &  0.55 &  0.37 &      0.44 &    1672 & (1664, 1680)\\
 StarNet (our) &  0.53 &  0.48 &      \textbf{0.50} &    1462 & (1430, 1497)\\
\bottomrule
\end{tabular}
\caption{Performance metrics on M2.
For probabilistic methods (StarNet and PCAT)
the ``\#stars" columns provide the posterior mean along with the 5th and 95th posterior percentiles
for the number of stars.
The number of stars in the ground-truth Hubble catalog is 1114. }
\label{tab:summary_stats}
\end{table}



\begin{figure}[tb]
    \centering
    \includegraphics[width=0.99\textwidth]{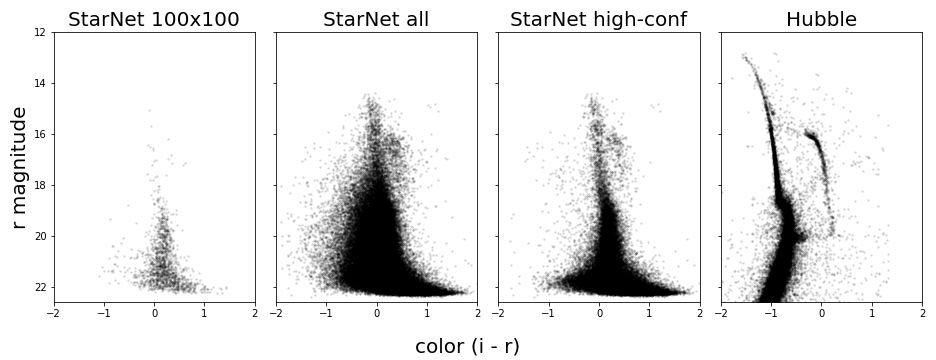}
    \caption{Color-magnitude diagrams from StarNet and Hubble. From left to right, color-magnitude diagrams constructed from: the same $100\times100$ subimage as was cataloged in \cite{Feder_2019}; 
    the StarNet catalog derived from the entire $1000 \times 1000$ image of M2; 
    that StarNet catalog, filtered to stars with posterior SD(flux) $< 1$; the Hubble catalog. 
}
    \label{fig:m2_cmd}
\end{figure}

In a subsequent experiment, 
we go beyond the $100\times100$ subimage
cataloged by \cite{Feder_2019} and 
catalog the entire M2 globular cluster 
contained in a $1000 \times 1000$-pixel image (Figure~\ref{fig:m2_region}). 
We produce a color-magnitude diagram on this entire region (Figure~\ref{fig:m2_cmd}). 
On the entire region, a second distinct cluster, 
shifted to the right in color, 
appears in addition to the main sequence of stars.
The second cluster becomes more apparent after we filter to high-confidence stars in the StarNet catalog, 
defined as stars with flux posterior standard deviation less than one. 
This second cluster, corresponding to a collection of red giants, 
is undetectable without the ability of StarNet to 
scale to larger images. 

However, 
the patterns are less definite in the StarNet color-magnitude diagram than in the Hubble color magnitude diagram.
There is more spread in the StarNet color estimates, particularly at faint magnitudes. 
This is due to the superior resolution of the Hubble telescope; near the heavily saturated core of the M2 cluster, stars are near impossible to deblend in the SDSS image, and our performance suffers (Appendix~\ref{sec:test_m2}).


\subsection{Results on the DECam Survey}
\label{sec:results_on_decam}

\begin{figure}[!htb]
    \centering
    \begin{subfigure}[T]{0.45\textwidth}
    \centering
    \includegraphics[width=\textwidth]{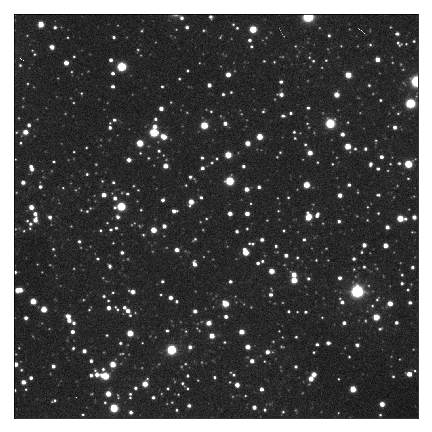}
    \end{subfigure}
    \hfill
    \begin{subfigure}[T]{0.5\textwidth}
    \centering
    \includegraphics[width=\textwidth]{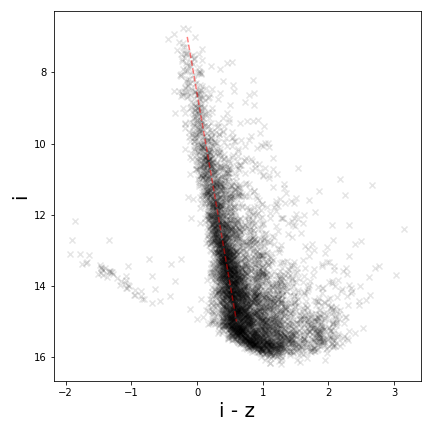}
    \end{subfigure}
    \caption{(Left) A 1000 x 1000 pixel subregion of the DECam survey. 
    (Right) Color magnitude diagram for the DECam image. Red dashed line highlights
    the inferred blue main-sequence stars}
    \label{fig:decaps}
\end{figure}

We next demostrate StarNet on a larger region of the sky.
The DECam survey imaged stars in our own Milky Way, and we chose
a $4000 \times 2000$ frame centered at coordinates $\text{RA} = 266.044^\circ$ and
$\text{DEC} = -28.88111^\circ$. See Figure~\ref{fig:decaps} for an example image.

The DECam image is somewhat sparser than M2.
Thus, we set the Poisson mean parameter of the star density lower smaller than on M2 to fifty stars per $100 \times 100$-pixel image.
This allowed us to use larger $10\times 10$-pixel tiles with $20\times20$-pixel
padded tiles. 
We produced
a catalog for the full $4000 \times 1000$ frame, consisting of $9,000$ stars. 
The color-magnitude diagram shows a sequence of blue
stars that are reddened at fainter magnitudes. 

\subsection{Runtime}
\label{sec:runtime}
We ran SGD to minimize the expected forward KL
for 400 epochs; at each epoch, 200 images of size $100\times100$ pixels were sampled from the generative model.
We performed optimization with Adam~\citep{kingma2014adam}.
On a single NVIDIA GeForce RTX 2080 Ti GPU,
this fitting procedure took one hour.

After fitting the variational posterior,
computing the approximate posterior
(that is, producing the distributional parameters of the variational approximation) given either the $1000\times1000$ M2 image
or the $4000 \times 2000$ DECam image
takes less than a second. 
By comparison, the reported runtime of PCAT, which uses MCMC, is 30 minutes on a $100 \times 100$ pixel image~\citep{Feder_2019}.

The speed at inference time (which excludes training time) gives StarNet the scaling characteristics necessary for processing large astronomical surveys.
A single SDSS image is $1489 \times 2048$ pixels.
Based on the reported 30-minute runtime of PCAT for a $100\times100$ pixel subimage, we project that
the runtime to process the full image would be $30\text{ min} \times 14 \times 20 = 8400$ minutes, or almost six days.
The SDSS survey consists of nearly one million images, and thus scaling PCAT to the entire SDSS survey would be infeasible.
The upcoming LSST survey will be 300 times larger than SDSS.

On the other hand, StarNet incurs a one-time cost to fit the variational distribution with synthetic data; this cost is then amortized over a potentially large region of the sky.
Re-fitting StarNet may nonetheless be necessary when the model parameters
such as the background or PSF change---which is the case for 
large ground-based astronomical surveys, where data is collected over many nights. 
The SDSS data processing pipeline, for instance, estimates a new PSF and background
for each new frame. 
Even assuming a new StarNet refit for each SDSS frame, 
StarNet is still $100\times$ faster than PCAT. 

We can further push the scalability of StarNet by 
amortizing over a range of model parameters such as the background and PSF. 
With appropriate priors on these model parameters,
fitting StarNet using the  
expected forward KL enables it to generalize across a diverse set of images 
and further reduce the need for retraining.

\section{Conclusion}
\label{sec:conclusion}

StarNet employs forward variational inference and is more accurate than both a recently published MCMC-based probabilistic cataloger and a widely used non-model-based
procedure.
In the framework of probabilistic modeling,
StarNet produces catalog uncertainties captured by a posterior over the set of all catalogs.
Importantly, unlike current MCMC approaches, StarNet also has the capacity to scale probabilistic cataloging to process large astronomical surveys.

The quality of StarNet detections is the result of optimizing the forward KL, a different objective than the one traditionally used in variational inference.
Optimizing the forward KL allows the variational posterior to be fit on large amounts of complete data (i.e., images along with their latent catalogs) generated from StarNet's statistical model.

While this work focuses on stars, our methodology can be extended to
include more general light sources, such as galaxies.
One promising direction is to incorporate a highly accurate deep generative model of galaxies~\citep{Regier2015ADG, Reiman_2019_gans_deblend, lanusse2020deep, Arcelin_2020} into the StarNet model.
The statistical framework in this research lays the foundation for building flexible models to incorporate the cataloging of other celestial objects.

Future astronomical surveys will produce far more data than past surveys.
As telescopes peer deeper into space, fields will reveal more sources and images will become more crowded.
The uncertainties in crowded fields necessitate a probabilistic approach.
Our method holds the promise of providing  a scalable inference tool that can meet the challenges of future surveys.


\section*{Acknowledgements}
RL acknowledges support from the NSF Graduate Research Fellowship Program.
JR acknowledges support for this work from the National Science Foundation (OAC-2209720) and the Department of Energy (DE-SC0023714).
This paper has been approved by the LSST Dark Energy Science Collaboration following an internal review. 
The internal reviewers were Bastien Arcelin, Fran\c{c}ois Lanusse, and Peter Melchior. 
The authors also thank Derek Hansen, Ismael Mendoza, and Zhe Zhao for providing thoughtful feedback about this manuscript.


\newpage

\appendix
\renewcommand\thefigure{A.\arabic{figure}}
\renewcommand\thetable{A.\arabic{table}}
\setcounter{figure}{0}
\setcounter{table}{0}

\section{Evaluating the Variational Distribution}
\label{sec:eval_var_distr}

Optimizing the expected forward KL requires evaluating
$q_\eta(z \mid x)$ for a given catalog
\[z = \{N, (\ell_i, f_{i,1}, ..., f_{i,B})_{i = 1}^N\}.\]
By~\eqref{eq:pull_back_of_q}, 
it suffices to evaluate $\tilde q_\eta(\tau^{-1}(z) \mid x)$, 
where $\tau$ is the mapping from tile latent variables $\tilde z$ to the catalog $z$ as described in Section~\ref{sec:factorization}, 
and $\tilde q_\eta$ is the distribution on tile latent variables. 

Here, $\tau^{-1}(z)$ is a {\itshape set} of tile latent variables because the mapping from tile latent variables $\tilde z$ to catalogs $z$ is not injective, as we now explain.  

Locations in the catalog $\{\ell_i\}_{i=1}^N$
determine the number of stars on tile $(s,t)$. 
The number of stars $\tilde N^{(s,t)}$ is simply the count of the locations that reside within that tile:
\begin{align}
\tilde N^{(s,t)} = \sum_{i=1}^N 
\mathbf 1 \Big\{\ell_i\in [Rs, R(s+1)] \times [Rt, R(t+1)]\Big\},
\end{align}
where $\mathbf{1}\{\cdot\}$ is the indicator function, equal to one its predicate is true if true and zero otherwise.

Now, consider $\tilde\ell^{(s, t)}$ and $\tilde f^{(s, t)}$, the triangular array of locations and fluxes on tile $(s,t)$. 
For each $(s,t)$, the $\tilde N^{(s,t)}$-th row 
of the triangular array of fluxes and locations is
determined by the locations and fluxes of stars imaged in tile $(s,t)$; they are determined by the catalog $z$. However, the other rows 
of the triangular arrays are not determined by 
the catalog $z$; they are free to take any value in their domain. Therefore, the mapping $\tau$ is not injective. 

Thus, evaluating the probability of $\tau^{-1}(z)$ under $\tilde q_\eta$ requires marginalizing over the rows of the triangular arrays $\ell^{(s, t)}$ and $\tilde f^{(s, t)}$ that are not determined by $z$. However, 
because $\tilde q_\eta$ fully factorizes, the terms 
where $n \not= \tilde N^{(s,t)}$ do not enter the
product
after marginalization.
On each tile $(s,t)$,
\begin{align}
    \tilde q_\eta\big(\tilde N^{(s, t)}, \tilde \ell^{(s, t)}, \tilde f^{(s, t)} \mid x\big) &= 
    \tilde q_\eta(\tilde N^{(s,t)} \mid x) 
    \prod_{n = 1}^{N_{max}}
    \tilde q_\eta\big((\tilde \ell_{n,i}^{(s, t)}, 
    \tilde f_{n,i}^{(s, t)})_{i=1}^n \mid x\big) \\
    &= \tilde q_\eta(\tilde N^{(s,t)} \mid x) 
    \tilde q_\eta\big((\tilde \ell_{n,i}^{(s, t)}, 
    \tilde f_{n,i}^{(s, t)})_{i=1}^n \mid x\big)\Bigg|_{n = \tilde N^{(s,t)}}.
\end{align}

In words, given a catalog $z$,
first convert $z$ to tile latent variables;
then on each tile, it suffices to evaluate $\tilde q_\eta$ only at the rows of triangular 
arrays determined by the number 
of stars falling in each tile. 

The last technical detail is computing the 
probability for 
a given row of a triangular array. 
Let $(\tilde \ell_i, \tilde f_i)_{i = 1}^n$ generically denote the tile latent variables in the $n$-th row of a triangular array, on some tile $(s,t)$.
Because catalogs are sets,
each entry $(\tilde \ell_i, \tilde f_i)$ in the catalog must 
be matched with corresponding variational parameters, 
and the probability of the set
$(\tilde \ell_i, \tilde f_i)_{i = 1}^n$ under $q$
is given by the sum over 
the permutations of the possible matches: 
\begin{align}
    q((\tilde \ell_{i}, \tilde f_{i})_{i = 1}^n | x) = \sum_\pi \Big\{ \prod_{i=1}^n \text{LogitNormal}(\tilde \ell_{\pi(i)}; \mu_{\ell_{i}}, \nu_{\ell_{i}})\times 
	\text{LogNormal}(\tilde f_{\pi(i)}; \mu_{f_{i}}, \sigma^2_{f_{i}})\Big\}
\end{align}
where the sum is taken over all permutations on $\{1, ..., n\}$. This is feasible because on each tile $N_{max} = 3$, so we only need to enumerate $3!=6$ possibilities. 
\section{Reparameterized and REINFORCE Gradients}
\label{sec:reparam_details}

The ELBO objective~\eqref{eq:elbo} is of the form
\begin{align}
    \mathcal{L}(\eta) = \Expect_{q_\eta(z)}[f_\eta(z)].
    \label{eq:gen_elbo}
\end{align}
The parameter $\eta$ is to be optimized, and $z$ is the latent variable. The integrating distribution $q$ and the function $f$ depend on $\eta$.

The REINFORCE estimator~\citep{Williams1992reinforce} is a general-purpose unbiased estimate for the gradient of~\eqref{eq:gen_elbo}.
It is given by
\begin{align}
    g_{\textrm{rf}}(z) = \nabla_\eta f_\eta(z) +
            f_\eta(z)  \nabla_\eta \log q_\eta(z)
    \quad \text{for }
    z\sim q_\eta(z).
\end{align}
The REINFORCE estimate is unbiased for the true gradient:
\begin{align}
    \Expect_{q_\eta(z)}[g_{\textrm{rf}}(z)] &=
    \int q_\eta(z) \nabla_\eta f_\eta(z) \; dz+
    \int q_\eta(z) f_\eta(z)  \nabla_\eta \log q_\eta(z)\; dz \\
    &= \int q_\eta(z) \nabla_\eta f_\eta(z) \; dz+
    \int f_\eta(z) \nabla_\eta q_\eta(z)  \; dz \\
    &= \int \nabla_\eta[q_\eta(z) f_\eta(z)] \; dz \\
    &= \nabla_\eta \int q_\eta(z) f_\eta(z) \; dz
    = \nabla_\eta \Expect_{q_\eta(z)}[f_\eta(z)],
\end{align}
assuming that $f$ is well-behaved so that integration and differentiation can be interchanged.

In many applications,
the REINFORCE estimator has too high variance to be useful. One way to lower the variance is to introduce a control variate $C$
\citep{ranganath2013black}, and estimate the gradient as 
\begin{align}
    g_{\textrm{cv}}(z) = \nabla_\eta f_\eta(z) +
        (f_\eta(z) - C)  \nabla_\eta \log q_\eta(z)
    \quad \text{for }
    z\sim q_\eta(z).
\end{align}
This estimate remains unbiased because the score function 
$\nabla_\eta \log q_\eta(z)$ is zero mean under $q$.

A simple but often effective choice of control variate is to let 
$C$ be a second evaluation of $f_\eta$ at an independently drawn $z'\sim q$:
\begin{align}
    g_{\textrm{cv}}(z) = \nabla_\eta f_\eta(z) +
        (f_\eta(z) - f_\eta(z')  \nabla_\eta \log q_\eta(z)
    \quad \text{for }
    z, z' \overset{\mathrm{iid}} \sim q_\eta.
    \label{eq:control_var}
\end{align}
This estimate is unbiased conditional on $z'$ and hence unconditionally unbiased as well. 
We use this control variate for our experiments involving 
the REINFORCE estimator in Section~\ref{sec:elbo_sleep_compare}.

Alternatively, the reparameterized gradient~\citep{rezende2014stochastic, kingma2013autoencoding} can be used when there exists some distribution $F$ not involving $\eta$ and a differentiable mapping $h_\eta$ such that
\begin{align}
    w \sim F \implies h_\eta(w) \sim q_\eta.
\end{align}
For example, if $q_{\eta}(z) = \mathcal{N}(z; 0, \eta)$ that is, a Gaussian with zero mean and variance $\eta$, one possibility is to let $F$ be the standard Gaussian and $h_\eta(w) = w \sqrt{\eta}$.
The gradient of $\mathcal{L}(\eta)$ can then be written as
\begin{align}
    \nabla_\eta \Expect_{q_\eta(z)}[f_\eta(z)] &=
        \nabla_\eta \Expect_{w\sim F}[f_\eta(h_\eta(w)] = \Expect_{w\sim F}[\nabla_\eta f_\eta(h_\eta(w)],
\end{align}
again assuming the interchangability of integrals and derivatives.
Unbiased gradients arise from the chain rule: 
\begin{align}
    g_{\textrm{rp}}
    = \nabla_\eta f_\eta(h_\eta(w))
    = \nabla_z f_\eta(z)\Big|_{z = h_\eta(w)}
    \nabla_\eta h_\eta(w) \quad \text{for } w\sim F.
\end{align}

The reparameterized gradient includes gradient information $\nabla_z f_\eta(z)$, while the REINFORCE gradient does not. Taking into account the structure of $f$ through its gradient lowers the variance of reparameterized gradient in comparison to the REINFORCE gradient.
However, if $z$ contains discrete components, there cannot be a differentiable mapping $h_\eta$, and the reparameterization trick will not apply.

\subsection{Gradients for the Empirical Comparison of KL Objectives}
In experiments of Section~\ref{sec:elbo_sleep_compare}, 
we used a combination of reparameterized and REINFORCE plus control variate gradients.
Let $\tilde N$ be the vector of per-tile number of stars (a discrete component) and $y$ be the locations and fluxes (continuous components)
Our variational distribution factorizes, so we write the expectation as
\begin{align}
 \mathcal{L}(\eta) = \Expect_{q_\eta(\tilde N)}\Expect_{q_\eta(y)}[f_\eta(\tilde N, y)].
 \label{eq:elbo_factorized}
\end{align}
We use the REINFORCE estimator with control variate \eqref{eq:control_var} for the outer expectation and the reparameterization trick for the inner expectation. 
We first apply REINFORCE to the outer expectation:
\begin{align}
    \nabla_\eta  &
    \Expect_{q_\eta(\tilde N)}\Expect_{q_\eta(y)}\Big[f_\eta(\tilde N, y)\Big] \\
    &=  \Expect_{q_\eta(\tilde N)}\Big[ \nabla_\eta \log q_\eta(\tilde N) \Expect_{q_\eta(y)}\big[f_\eta(\tilde N, y) - 
    \Expect_{q_\eta(\tilde N)}[f_\eta(\tilde N, y)]\big] +
    \nabla_\eta \Expect_{q_\eta(y)}[f_\eta(\tilde N, y)] \Big]\\
    &\approx \nabla_\eta \log q_\eta(\tilde N) \Expect_{q_\eta(y)}[f_\eta(\tilde N, y) - f_\eta(\tilde M, y)] +
    \nabla_\eta \Expect_{q_\eta(y)}[f_\eta(\tilde N, y)]
    \label{eq:reinforce_partial}
\end{align}
 for $\tilde N,\tilde M \overset{iid}\sim q_\eta$.
Then we use the reparameterization trick for $y$, so
\begin{align}
    \Expect_{q_\eta(y)}[f_\eta(\tilde N, y) - f_\eta( \tilde M, y)] &\approx f_\eta(\tilde N, h_\eta(w)) - f_\eta(\tilde M, h_\eta(w))\\
    \nabla_\eta \Expect_{q_\eta(y)}[f_\eta(\tilde N, y)] &\approx  \nabla_y f_\eta(\tilde N, y)\Big|_{y = h_\eta(w)}
    \nabla_\eta h_\eta(w)
    \label{eq:reparam_partial}
\end{align}
for $w \sim F$, where $h_\eta$ and $F$ are chosen appropriately. Combining~\eqref{eq:reinforce_partial} and~\eqref{eq:reparam_partial}, our gradient estimator is
\begin{align}
    g(z) = \nabla_\eta \log q_\eta(\tilde N)
    [f_\eta(\tilde N, h_\eta(w)) - f_\eta(\tilde M, h_\eta(w))] +
    \nabla_y f_\eta(\tilde N, y)\Big|_{y = h_\eta(w)}
    \nabla_\eta h_\eta(w).
    \label{eq:reiforce_final}
\end{align}
Equation~\eqref{eq:reiforce_final} is what our main text called the ``REINFORCE gradient."
These gradients produced the optimization path in Figure~\ref{fig:optim_path}(a).

The ``reparameterized" gradient in Section~\ref{sec:elbo_sleep_compare} requires integrating out $\tilde N$.
Here, we write the outer expectation in~\eqref{eq:elbo_factorized} as a summation of $4^{N_{max}+1}$ terms
(recall our experiments have four tiles, and at most $N_{max}$ stars per tile), 
with each term representing a different possible 
assignment of the number of stars to each tile: 
\begin{align}
\mathcal{L}(\eta) = \sum_{\tilde n}\Expect_{q_\eta(y)}[f_\eta(\tilde n, y)].
\end{align}
Then, the reparameterization trick is applied to each term of the summation.
Stochastic gradients are computed as,
\begin{align}
g(z) = \sum_{\tilde n=1}\nabla_y f_\eta(\tilde n, y)\Big|_{y = h_\eta(w)}
\nabla_\eta h_\eta(w) \quad \text{for } w\sim F.
\end{align}
Gradients of this form produced the optimization path in Figure~\ref{fig:optim_path}(b).
No REINFORCE estimates were required. 

Notice that to compute these re-parameterized gradients we require 
a summation of $(S \times T)^{N_{max}+1}$ terms 
(recall from the main text that $S \times T$ is the total number 
of tiles in an image). 
For large images, i.e. those requiring many tiles, computing re-parameterized gradients would be infeasible. 


\section{Experiments on Synthetic Data}

We present results on a set of synthetic data experiments.
We first demonstrate the ability of StarNet to deblend two simulated stars. 
Next, we revisit the coverage of StarNet credible intervals. 
Finally, we empirically demonstrate the effect of padded tiles
in our architecture. 

\subsection{Deblending Two Stars}

\begin{figure}[tb]
    \centering
    \begin{subfigure}{0.8\textwidth}
        \includegraphics[width=\textwidth]{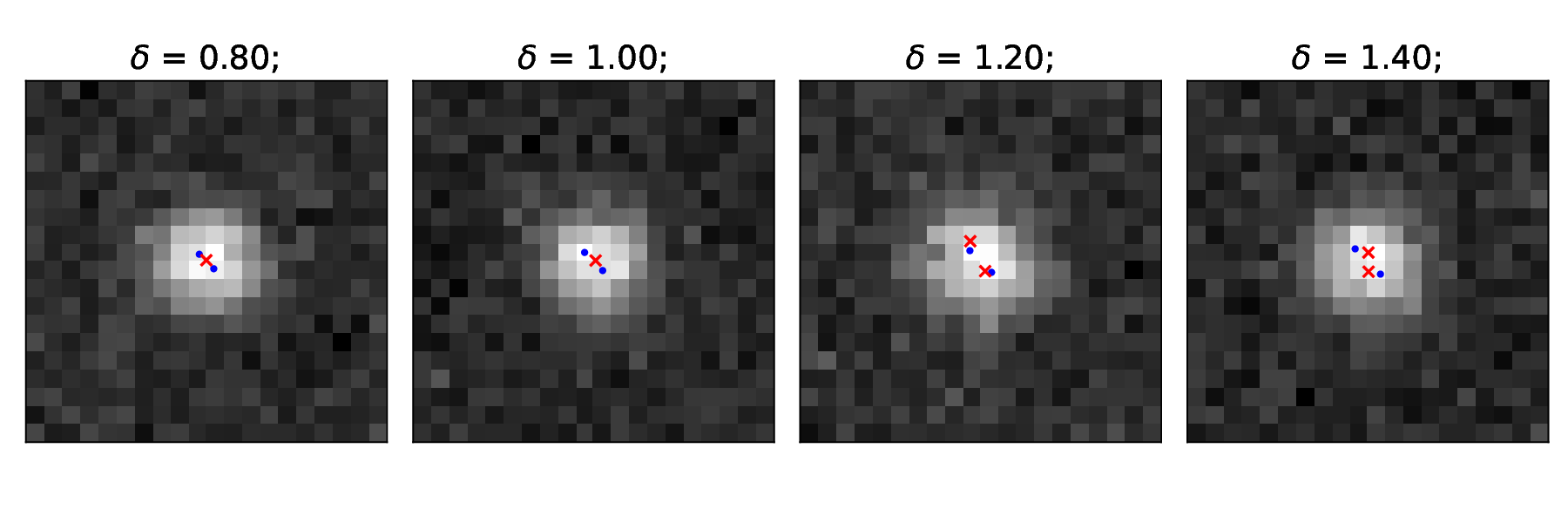}
    \end{subfigure}
    \begin{subfigure}{0.8\textwidth}
        \includegraphics[width=\textwidth]{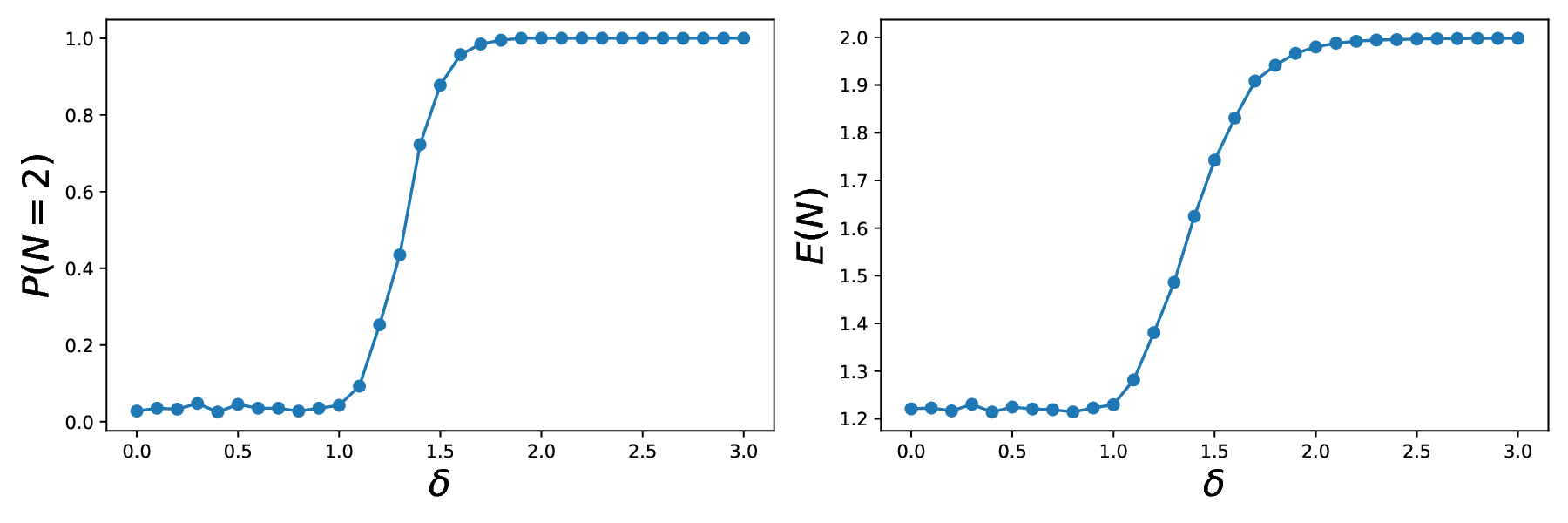}
    \end{subfigure}
    \caption{(Top row) Simulated images with two stars separated by distance $\delta$ in pixels.
    True locations are in blue. StarNet MAP locations in red. 
    In these examples, the StarNet MAP catalog correctly contains two stars when separated by $\delta \geq 1.2$,
    but only estimated one star when $\delta \leq 1$.
    (Bottom left) The probability that $N$, the number of sources in the image, equals two
    under the variational posterior, as $\delta$ varies.
    (Bottom right) The expectation of $N$ under the variational posterior as $\delta$ varies. }
    \label{fig:deblending_fig}
\end{figure}

We set up an experiment to study the
ability of StarNet to deblend two simulated stars.
On a $20\times20$ pixel image,
we simulate two stars of equal flux at distance $\delta$ pixels apart, and
examine the approximate posterior produced by StarNet (Figure~\ref{fig:deblending_fig}).
We generate the stars with the DECam PSF, which 
has a full width at half maximum 
(defined as the diameter at which the PSF is half its peak brightness) of 4.2 pixels. 
The threshold of near-perfect deblending is a distance of $\delta = 1.5$ pixels, which is less than half the PSF 
full width at half maximum. 

\subsection{Coverage of Credible Intervals}
\label{sec:coverage}

We have seen that on M2, the StarNet 95\% posterior interval did not contain the ground truth number of stars (Section~\ref{sec:results_on_m2}).
On M2, we attribute this to model mis-specification, specifically due to an imperfectly
estimated background.

\begin{figure}[tb]
    \centering
    \includegraphics[width=0.45\textwidth]{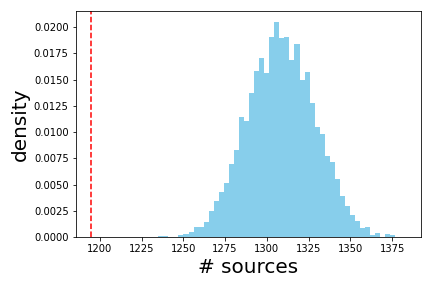}
    \caption{Distribution of 5000 samples of the number of sources from the StarNet approximate posterior.
    The true number of sources demarcated in red. }
    \label{fig:starnet_density}
\end{figure}

We check the coverage of StarNet posterior intervals on synthetic data
to reveal any issues other than model mis-specification that may explain these results.
We sample a single $100\times100$ pixel image from the generative model. 
On this sampled image, the true number of stars, $N = 1195$,
is still considerably smaller than the 0.01-th percentile of
the approximate posterior distribution (Figure~\ref{fig:starnet_density}).

We attribute this over-estimation to the spatial independence of the approximate posterior.
Specifically, StarNet \textit{overestimates} the number of sources close to tile boundaries.
Heuristically, for a
a source located in the interior of the tile
within $\epsilon$ of a boundary (but still far from a corner),
StarNet should assign a probability of $\frac{1}{2}$ for
having one source, and a probability $\frac{1}{2}$ for having none, as $\epsilon \rightarrow 0$.
Should this be the case, then over the entire image, which consist of many tiles,
a source on a tile boundary is correctly accounted for: it has a 50-50 chance of being assigned to one tile or another
in the approximate posterior, and this source contributes 
a count of one to the posterior expectation on the number of stars. 

\begin{figure}[tb]
    \centering
    \includegraphics[width=0.8\textwidth]{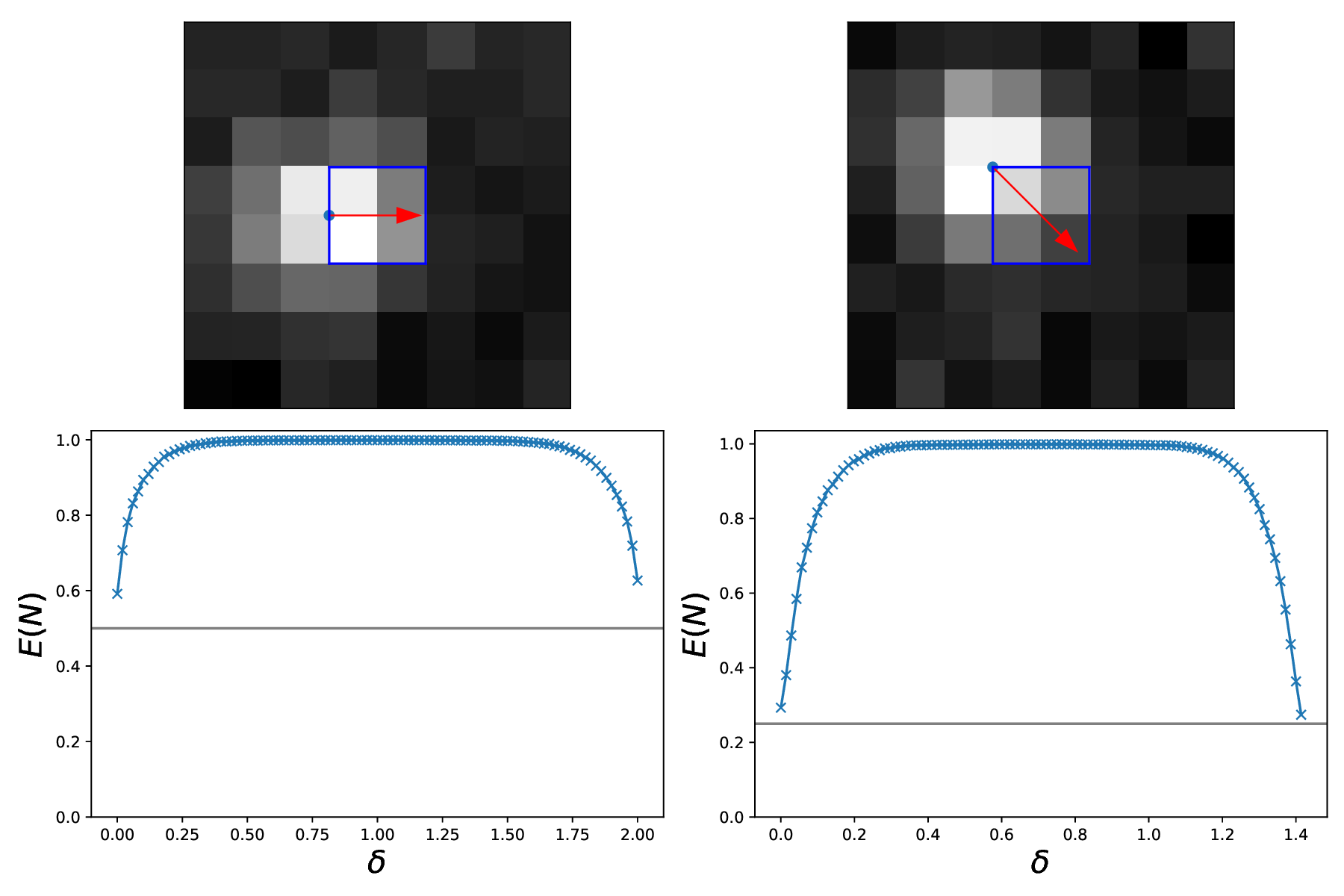}
    \caption{The expected number of sources under the StarNet approximate posterior as a function of distance from the 
    tile edge (in pixels).
    On the left, we place a star on the left-most edge, and move its location $\delta$ pixels to the right.
    On the right, we place a star on the top-left corner, and move its location $\delta$ pixels towards the bottom-right corner.}
    \label{fig:starnet_edges}
\end{figure}

However, we observe empirically that as a source approaches the edge of a tile,
the posterior probability that $N = 1$ approaches a number
slightly \textit{larger} than $\frac{1}{2}$ (Figure~\ref{fig:starnet_edges}).
Therefore, the approximate posterior overestimates the expected number of sources in the full image.

\begin{figure}[tb]
    \centering
    \includegraphics[width=0.5\textwidth]{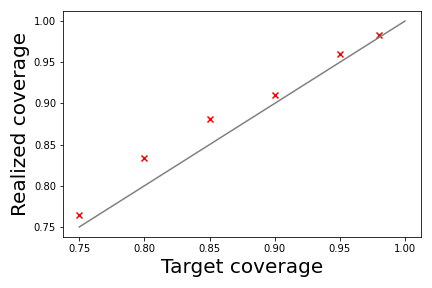}
    \caption{Coverage test in simulated $100\times100$ images where all true stars are constrained to be 0.1-pixels away from any tile
    boundary.
    We simulate 1000 images from the generative model, and for each image, we compute a $(1 - \alpha)$-level posterior interval for the number of stars by taking the $\frac{\alpha}{2}$ and $(1 -\frac{\alpha}{2})$-th percentiles of 5000 StarNet posterior samples.
    For each $\alpha$, we plot the observed coverage against the target $(1 - \alpha)$-level coverage. }
    \label{fig:coverage_good}
\end{figure}

To illustrate the effect of tiles, we simulate images with the constraint that all sources are at least 0.1-pixels from all tile edges. 
In this case, then the StarNet approximate posterior has much closer to correct coverage (Figure~\ref{fig:coverage_good})

\subsection{Effects of Tile Padding}
\begin{figure}[tb]
    \centering
    \includegraphics[width=0.8\textwidth]{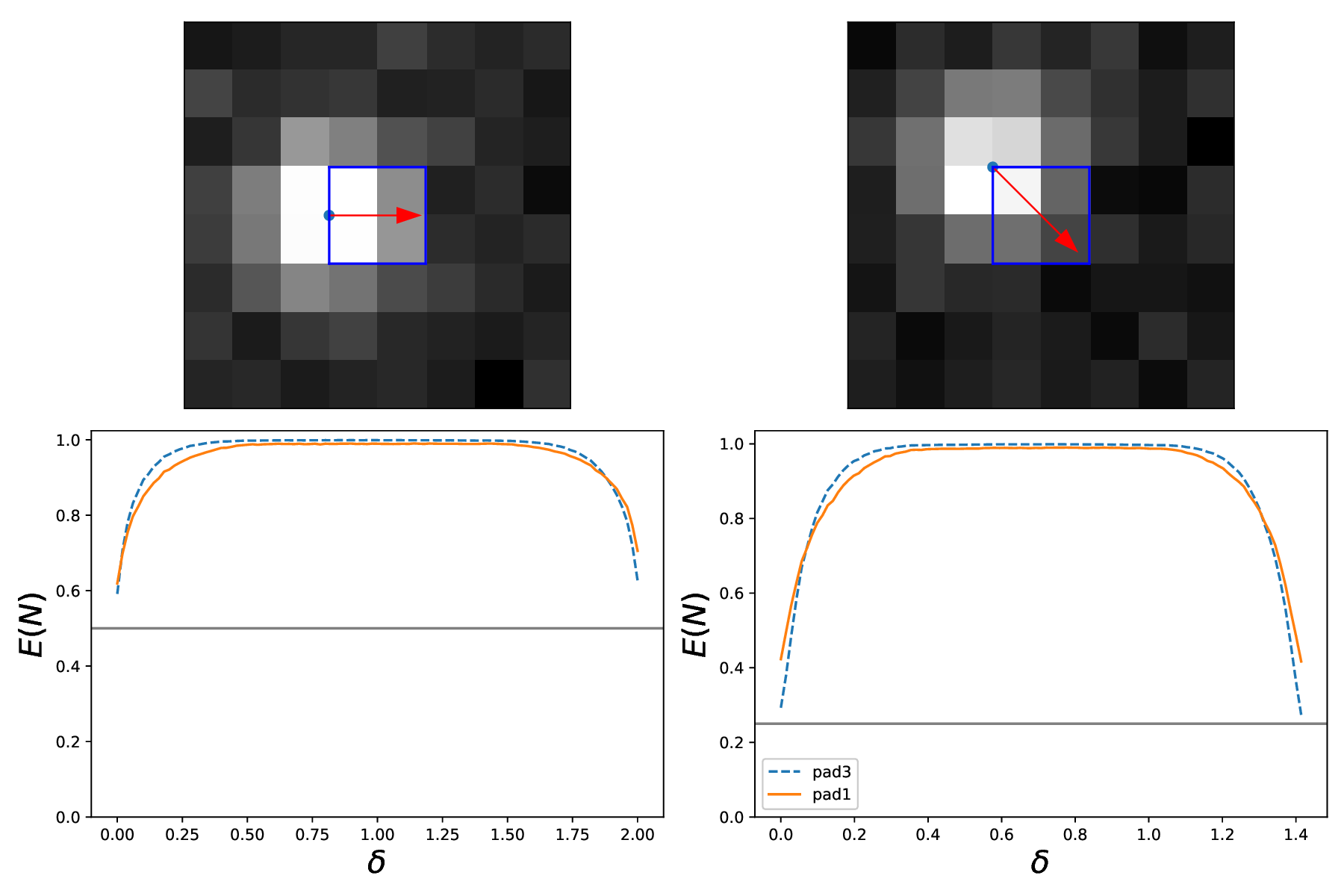}
    \caption{The same experimental setup as Figure~\ref{fig:starnet_edges}, but we 
    study the effect of padding the neural network input tiles. 
    In dashed blue, the posterior when 
    StarNet uses three pixels of padding. 
    In solid orange, the posterior when only one pixel of padding is used.
    }
    \label{fig:starnet_edges_lesspad}
\end{figure}

We study the effect of padding the input 
tiles to the StarNet neural network architecture. 
We re-visit the investigation of tile boundary effects 
described in the subsection above, where 
we simulate a source closer and closer to the tile boundary. 
The experiments above used the M2 network, 
which uses $2\times 2$ tiles and three pixels of padding. 
With only one pixel of padding, sources become even more 
over-estimated on tile edges (Figure~\ref{fig:starnet_edges_lesspad}). 

\section{Sensitivity to Prior Parameters}
\label{sec:prior_sensitivity}

We examine the sensitivity of StarNet to prior parameters $\mu$ and $\alpha$ on the image M2. 
Recall $\mu$ is the prior mean number of stars per pixel~\eqref{eq:n_prior};
$\alpha$ is the power law slope on the $r$-band fluxes~\eqref{eq:flux_prior}. 
In the results of Section~\ref{sec:results_on_m2}, $\mu=0.12$ and  $\alpha = 0.5$. 

The model is robust to these prior choices. 
In fact, the variation in performance metrics
due to prior choices is about the same as the 
variation due to random refits.
Thus, for these prior sensitivity experiments, 
we initialize the neural network weights
using the network fit at the original prior. 

Over a  range of $\mu$ between 0.08 (corresponding to 
an prior average of 800 stars on a $100\times100$ image) 
and 0.20 (corresponding to 2000 stars)
the F1-score remains steady between 0.49 and 0.51 (Table~\ref{tab:mu_sensitivity}). 
Similar robustness in F1 hold when $\alpha$ varies between 
0.25 and 1.0 (Table~\ref{tab:alpha_sensitivity}). 

\begin{table}[!tb]
\centering
\begin{tabular}{l|ccc}
\toprule
$\mu$ &   TPR &   PPV &  F1 score \\
\midrule
0.08 &  0.47 &  0.53 & 0.50 \\
0.10 &  0.49 &  0.53 & 0.51 \\
0.12 &  0.53 &  0.48 & 0.50 \\
0.14 &  0.50 &  0.46 & 0.48 \\
0.16 &  0.51 &  0.48 & 0.50 \\
0.20 &  0.51 &  0.48 & 0.49 \\
\bottomrule
\end{tabular}
\caption{Performance metrics on M2 as a function of 
the prior parameter $\mu$. 
The smallest $\mu$, $\mu = 0.08$ corresponds to 
an prior average of 800 stars on a $100\times100$ image, 
while the largest, $\mu = 0.2$, corresponds to 2000 stars. }
\label{tab:mu_sensitivity}
\end{table}

\begin{table}[!tb]
\centering
\begin{tabular}{l|ccc}
\toprule
$\alpha$ &   TPR &   PPV &  F1 score \\
\midrule
0.25 &  0.47 &  0.54 & 0.51 \\
0.50 &  0.53 &  0.48 & 0.50 \\
0.75 &  0.56 &  0.45 & 0.50 \\
1.00 &  0.55 &  0.47 & 0.51 \\
\bottomrule
\end{tabular}
\caption{Performance metrics on M2 as a function of 
the prior parameter $\alpha$. }
\label{tab:alpha_sensitivity}
\end{table}

\section{Sensitivity to Refits}
\label{sec:refits}

Our optimization procedure uses stochastic optimization. 
We evaluate the sensitivity of our StarNet M2 performance metrics to ten random refits (Table~\ref{tab:refits}). 
The F1 score over the refits range
between 50\% and 55\%. Our comparisons 
to the performance of other methods, PCAT and DAOPHOT, 
are unaffected by random reruns. 

\begin{table}
\centering
\begin{tabular}{lrrr}
\toprule
 &       TPR &       PPV &        F1 \\
\midrule
0 &  0.53 &  0.48 &  0.50 \\
1 &  0.52 &  0.52 &  0.52 \\
2 &  0.53 &  0.57 &  0.55 \\
3 &  0.53 &  0.49 &  0.51 \\
4 &  0.53 &  0.52 &  0.52 \\
5 &  0.53 &  0.53 &  0.53 \\
6 &  0.55 &  0.54 &  0.55 \\
 7 &  0.51 &  0.54 &  0.52 \\
8 &  0.53 &  0.52 &  0.53 \\
9 &  0.51 &  0.55 &  0.53 \\
\bottomrule
\end{tabular}
\caption{Performance metrics on M2 for ten random refits of StarNet. }
\label{tab:refits}
\end{table}

\section{Other M2 Subregions}
\label{sec:test_m2}

\begin{figure}[tb]
    \centering
    \includegraphics[width=0.6\textwidth]{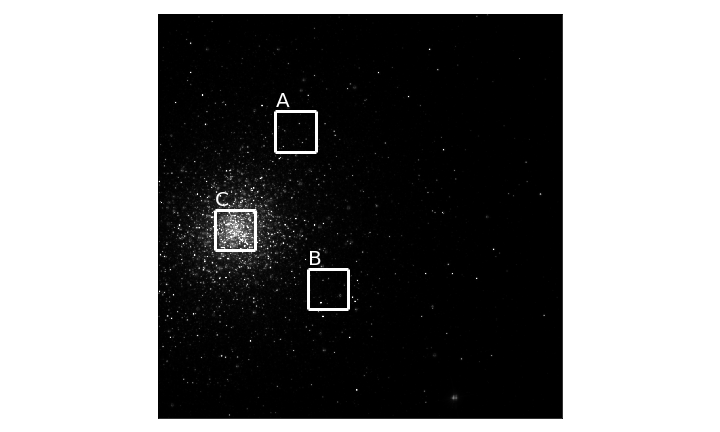}
    \caption{Subregions of M2 for the performance metrics in Table~\ref{tab:summary_stats_m2test}. The subregion labeled "A" was cataloged in the main text. 
}
    \label{fig:m2_test_regions}
\end{figure}

The initial $100\times100$ subregion of M2 considered in our main paper was located at pixel coordinates (630, 310) in SDSS run 2583, field 136, camera column 6. 
We evaluate StarNet on two other subregions of M2.
The first is another $100\times100$ pixel subregion of similar density as the original;
the second is in the center of globular cluster
(Figure~\ref{fig:m2_test_regions}).

The performance metrics on the center of the globular cluster 
suggest that deblending in this region is nearly impossible --- there are $15,000$ stars in this $100\times100$ subregion, 
averaging to more than one star per pixel
and is 
ten times as dense as the region considered in the main text.

In all the considered regions, our comparison with DAOPHOT is unchanged, and we continue to outperform DAOPHOT in F1 score (Table~\ref{tab:summary_stats_m2test}). 

\begin{table}[ht]
\centering
\begin{tabular}{ll|ccc|cc|c}
\toprule
     Region & Method &   TPR &   PPV &  F1 score &  \#stars & (q-5\%, q-95\%) & True \#stars \\
\midrule
 (A) &   DAOPHOT & 0.20 &  0.65 &      0.31 &     357 & --    & 1114\\
 (A) &  StarNet & 0.53 &  0.48 & 0.50 &    1462 & (1430, 1497)& 1114\\\hline
 (B) &   DAOPHOT & 0.21 &  0.65 &      0.32 &     310 & --    & 941\\
 (B) &  StarNet & 0.56 &  0.44 & 0.49 &    1384 & (1352, 1416)& 941\\\hline
 (C) &   DAOPHOT & 0.003 &  0.19 &      0.007 &     293 & --    & 15094\\
 (C) &  StarNet & 0.08 &  0.31 & 0.13 &    3306 & (3258, 3355)& 15094\\
 \bottomrule
\end{tabular}
\caption{Performance metrics of StarNet and DAOPHOT on the three regions labeled ``A", ``B", and ``C" in Figure~\ref{fig:m2_test_regions}.}
\label{tab:summary_stats_m2test}
\end{table}

\section{Neural Network Architecture Details}
\label{sec:supp_nn_architecture}

\begin{figure}[!tb]
    \centering
    \includegraphics[width=0.8\textwidth]{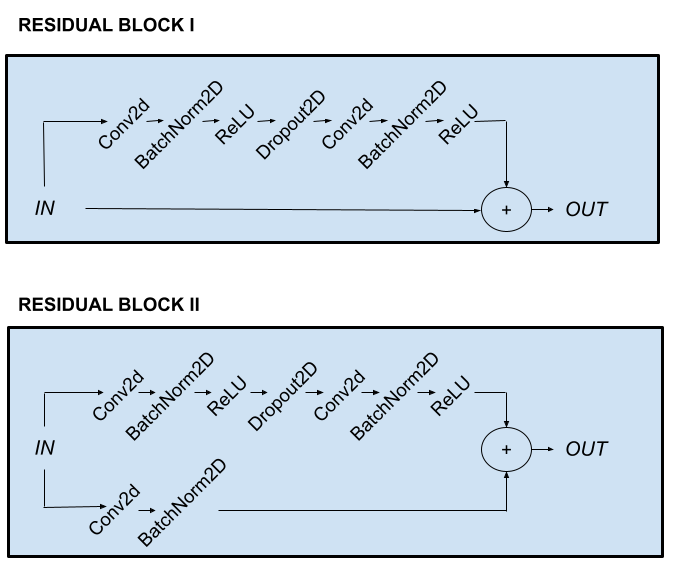}
    \caption{Details of the residual network blocks from Figure~\ref{fig:starnet_arch}.
    }
    \label{fig:conv_blocks}
\end{figure}

We detail the neural network architecture. 
Figure~\ref{fig:starnet_arch} shows a schematic of the architecture, and 
Figure~\ref{fig:conv_blocks} depicts specifically 
the residual network blocks. 

The first convolutional layer (green block, Figure~\ref{fig:starnet_arch}) has 17 output-channels, a kernel size of three, a stride of one, and one pixel of padding.
All convolutional layers inside residual block 1, as well as the convolutional layers on the top row of residual block 2 (Figure~\ref{fig:conv_blocks}) also have the same parameters. 
Only the convolutional layers on the bottom row of residual block 2 are different: they still have output channels of dimension 17, 
but down-sample using a kernel size of one, and a stride of 2. 
Inside the residual blocks, the dropout layers have dropout probability of 0.11399. 

The final fully connected block (red block, Figure~\ref{fig:starnet_arch}) has latent dimension 185, and a dropout probability of 0.013123.

\clearpage

\vskip 0.2in
\bibliography{bibliography}

\end{document}